\documentstyle[seceq,preprint,epsbox,tables]{jpsj}

\def\runtitle
{
Many Body Effects on Electron Tunneling through 
Quantum Dots in an AB Circuit
}

\def\runauthor
{
Wataru {\sc Izumida}
Osamu {\sc Sakai}
and Yukihiro {\sc Shimizu}
}

\title
{
{\small Many Body Effects on Electron Tunneling through
Quantum Dots in an AB Circuit}\\
\vspace{10ex}
Many Body Effects on Electron Tunneling through \\
Quantum Dots in an Aharonov-Bohm Circuit
}

\author
{
Wataru {\sc Izumida},\footnote{izumida@cmpt01.phys.tohoku.ac.jp}
Osamu {\sc Sakai}
and Yukihiro {\sc Shimizu}$^{1}$
}

\inst
{
Department of Physics, Tohoku University, Sendai 980-77,Japan\\
$^1$Department of Applied Physics, Tohoku University, Sendai 980-77,Japan
}

\recdate
{
October 15, 1996}

\abst
{
Tunneling conductance of an Aharonov-Bohm circuit including two quantum dots
is calculated
based on the general expression of the conductance
in the linear response regime of the bias voltage.
The calculation is performed in a wide temperature range
by using numerical renormalization group method.
Various types of AB oscillations appear
depending on the temperature and the potential depth of the dots.
Especially,
AB oscillations have strong higher harmonics components as a function of the magnetic flux
when the potential of the dots is deep.
This is related to the crossover of the
spin state due to the Kondo effect on quantum dots.
When the temperature rises up,
the amplitude of the AB oscillations becomes smaller 
reflecting the breaking of the coherency.
}

\kword
{
quantum dots,
Aharonov-Bohm effect,
Kondo effect,
numerical renormalization group method
}

\begin{document}
\sloppy
\maketitle

\clearpage

%
\section{Introduction}
\label{intro}

There are many systems that electron-electron interactions play
important roles in the physical properties of them.
A system of metal with dilute magnetic impurities 
is a good example.
Various anomalous phenomena
are observed at low temperatures,
related to the Kondo effect,
which is caused by 
competition between quantum coherence effects and electron-electron
interactions.~\cite{rf:Kondo_effect}

In the phenomenon of the electron tunneling
through 
a quantum dot between leads,
the effects of the electron-electron interactions in the dot 
have been investigated extensively.~\cite{rf:Kondo_dot_Ng,rf:Kondo_dot_JETP,rf:Kondo_dot_Kawa,rf:Oguri1,rf:Kondo_dot_Neq1,rf:Kondo_dot_Neq2,rf:Kondo_dot_Neq3,rf:Kondo_dot_Neq4,rf:And_dot_Eq0,rf:And_dot_Eq1,rf:And_dot_Neq2}
The importance of the Kondo effect in a quantum dot 
has been pointed out 
by some theoreticians.~\cite{rf:Kondo_dot_Ng,rf:Kondo_dot_JETP,rf:Kondo_dot_Kawa,rf:Oguri1,rf:Kondo_dot_Neq1,rf:Kondo_dot_Neq2,rf:Kondo_dot_Neq3,rf:Kondo_dot_Neq4}
The `Kondo effect in a quantum dot' 
will give the following features.
%
In low temperatures than the Kondo temperature of the system,
$T_{\rm K}$,
the conductance will show the coherent resonant tunneling 
caused by the Kondo resonance peak in the quantum dot.
When the temperature rises up than $T_{\rm K}$,
the Kondo resonance peak vanishes and the local spin freedom appears,
and so the electron tunneling
with inelastic spin scattering process will dominate.
Therefore the coherent resonant tunneling process will be broken.

The interference phenomenon
is the good indication for the coherency of the electronic state.
The Aharonov-Bohm (AB) circuits
with quantum dots have been investigated 
to study the coherency on quantum dots.~\cite{rf:AB_dot_Akera,rf:AB_dot_Yacoby,rf:AB_dot_Yacoby2,rf:AB_dot_Yeyati,rf:AB_dot_Hachen,rf:AB_dot_Bruder}
Akera had studied the system 
in which two quantum dots are included in AB circuit,
and shown that the amplitude of the AB oscillations depends on the spin state of the dots.~\cite{rf:AB_dot_Akera}
However,
his investigation was restricted only to the high temperature region than $T_{\rm K}$.
Main purpose of the present work 
is to study how the Kondo effect affects the AB oscillations.
We calculate 
the temperature dependence of the AB oscillations
in a wide temperature range including the crossover temperature
$T_{\rm K}$.

\begin{figure}[htb]
 \begin{center}
  \epsfile{file=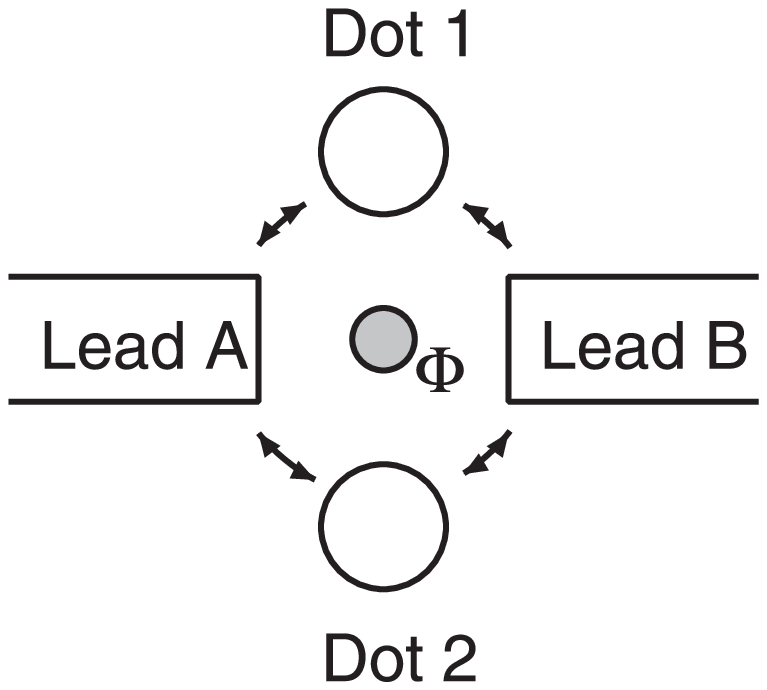,scale=.6}
  \caption
  {
   The geometry of the two quantum dots and leads.
   Electron tunneling from the one lead to the other lead 
   can occur only via quantum dots.
   The magnetic flux $\Phi$ is applied between leads and dots
   to cause AB interference effect.
  }
  \label{fig:system}
 \end{center}
\end{figure}

In this paper,
we consider the system shown in Fig. {\ref{fig:system}},
which is the same system that Akera studied.
The essential physics of this system can be given by 
the two impurity Anderson Hamiltonian.
Because our interests are related to the change of the coherency of the electronic state,
we have to calculate the conductance 
without assumptions on tunneling processes.
In other words,
the assumption,
such as the restriction to the `coherent resonant tunneling process'
or to the `sequential tunneling process',
should not be used.
Numerical Renormalization Group (NRG) method 
has been known as a useful method 
to study the many body systems
written by the Anderson Hamiltonian.~\cite{rf:NRG_Kri,rf:NRG_sakai,rf:NRG_shimizu}
This method has reliability 
in the temperature range 
near and below $T_{\rm K}$.
In this paper,
we develop this method  
to calculate the conductance without assumptions on tunneling processes.

Main results of this paper are as follows:
When the temperature rises up,
the amplitude of the AB oscillations becomes smaller
reflecting breaking of the coherency by spin inelastic scattering;
Various types of AB oscillations appear due to many body effects 
when the depth of dot's potential is changed;
AB oscillations have strong higher harmonics components 
as a function of the magnetic flux
reflecting the crossover of the spin state.

In section {\ref{model}},
the model Hamiltonian of the system is given 
and is transformed for numerical calculation.
The conductance formula is derived based on the linear response theory.
At the same time,
for comparisons,
the conductance formula based on the coherent resonant tunneling process
is derived by using the Friedel sum rule.
In section {\ref{cal}},
numerical results of the conductance are shown.
In section {\ref{sum}},
summary and discussion of this study are given.

%

%
\section{Formulation}
\label{model}

\subsection{Model and transformation}

For simplicity,
we consider following situation for our system in Fig. {\ref{fig:system}}.
Energy separations
between states of the single-electron orbits in a dot
are much larger than 
the other energy scales:
the Coulomb repulsion energy $U$,
temperature $T$,
and the level width $\Delta$.
This situation corresponds to the 
assumption that only one single-electron orbit in the dot
contributes to the current.
Two electrons can occupy in this orbit because of the spin degeneracy,
but the second electron need excess energy $U$ due to 
the Coulomb interaction between two electrons.
The orbit in the dot couples to those in leads,
and this causes the electron tunneling.
Tunneling from the one lead to the other lead 
can occur only via quantum dots.
The essential properties of the system can be written by 
the Anderson model as follows;
\begin{eqnarray}
H & = & H_{{\rm l}}+H_{{\rm d}}+ H_{{\rm l-d}}, 
\label{eq:totalH}\\
H_{{\rm l}} & = & \sum_{ \vec{k} \sigma} \varepsilon_{a \vec{k} \sigma}
                  a_{\vec{k} \sigma}^{\dagger} a_{\vec{k} \sigma}
                 +\sum_{ \vec{p} \sigma} \varepsilon_{b \vec{p} \sigma}
                  b_{ \vec{p} \sigma}^{\dagger} b_{ \vec{p} \sigma},
\label{eq:H_l}\\
H_{{\rm d}} & = & 
                \sum_{i=1,2 \hspace{1mm} \sigma} 
                \varepsilon_{{\rm d},i}
                d_{i \sigma}^{\dagger} d_{i \sigma}
                + \sum_{i=1,2} 
                U_{i} d_{i \uparrow}^{\dagger} d_{i \uparrow} 
                      d_{i \downarrow}^{\dagger} d_{i \downarrow},
\\
H_{{\rm l-d}} & = &  \sum_{ \vec{k} \sigma}
                     (v_{a \vec{k} , 1} 
                      e^{{\rm i} \frac{\varphi}{4}} d_{1 \sigma}^{\dagger} a_{ \vec{k} \sigma}
                    + v_{a \vec{k} , 2} 
                      e^{{\rm -i} \frac{\varphi}{4}} d_{2 \sigma}^{\dagger} a_{ \vec{k} \sigma}
                    +h.c.) \nonumber \\
                 && + \sum_{ \vec{p} \sigma}
                     (v_{b \vec{p} , 1} 
                      e^{{\rm -i} \frac{\varphi}{4}} d_{1 \sigma}^{\dagger} b_{ \vec{p} \sigma}
                    + v_{b \vec{p} , 2} 
                      e^{{\rm i} \frac{\varphi}{4}} d_{2 \sigma}^{\dagger} b_{ \vec{p} \sigma}
                    +h.c.), \nonumber \\
&& \label{eq:H_l-d}
\end{eqnarray}
where 
$a_{\vec{k} \sigma}$ 
is the annihilation operator of the electron 
with spin $\sigma$ 
and the state $\vec{k}$ in the left lead A,
$b_{\vec{p} \sigma}$ is that of in the right lead B,
and 
$d_{i,\sigma}$ 
is that of on the orbit in the $i$-th dot,
respectively.
The quantity $\varepsilon_{{\rm d},i}$ 
is the single-electron level of orbit in the $i$-th dot,
and is the effective depth of dot's potential.
The quantity $U_{i}$ is the Coulomb repulsion energy in the $i$-th dot. 
The terms $H_{{\rm l}}$ and $H_{{\rm d}}$ denote the leads and the dots,
respectively,
and 
$H_{{\rm l-d}}$ is the electron tunneling term between leads and dots.
The factor $e^{\pm {\rm i} \frac{\varphi}{4}}$ 
is due to the magnetic flux.
($\varphi = 2 \pi \Phi / \Phi_0$, 
where $\Phi$ is the magnetic flux enclosed by the circuit,
and $\Phi_0$ is the magnetic flux quantum,
$hc/e$.)

We transform the model Hamiltonian to fit numerical calculation.
We consider only the situation that
$\varepsilon_{\rm d,1} = \varepsilon_{\rm d,2} \equiv \varepsilon_{\rm d}$ and
$U_{1} = U_{2} \equiv U$ for simplicity.
Furthermore,
we assume that the geometry of the leads 
has symmetry with respect to the interchange of dot 1 and dot 2.
Moreover we also assume the symmetry 
with respect to the interchange of lead A and lead B.
Electron motion in the leads are restricted only 
to the one-dimensional motion along the lead.
(The state index $\vec{k}$ is reduced to $k$.)
Neglecting the $k$ dependence of the tunneling matrix,
we put the relations,
$v_{a k , 1} = v_{b p , 1} = v_{a k , 2} = v_{b p , 2} \equiv V$.
Finally, 
each term is rewritten as
\begin{eqnarray}
    H_{{\rm l}} 
    & = & 
    \sum_{k\sigma} \varepsilon_{k} 
    (\alpha_{{\rm s},k\sigma}^{\dagger} \alpha_{{\rm s},k\sigma}
    + \alpha_{{\rm a},k\sigma}^{\dagger} \alpha_{{\rm a},k\sigma}),
\label{eq:H_l_mod} \\
    H_{{\rm d}} & = & \varepsilon_{{\rm d}} (\sum_{\sigma} n_{{\rm e},\sigma}
    + \sum_{\sigma} n_{{\rm o},\sigma}) \nonumber \\
    && +  \frac{U}{2}(n_{{\rm e},\uparrow} n_{{\rm e},\downarrow}
    + n_{{\rm o},\uparrow} n_{{\rm o},\downarrow}) \nonumber \\
    && +  \frac{U}{4} (n_{{\rm e},\uparrow} + n_{{\rm e},\downarrow})
    (n_{{\rm o},\uparrow} + n_{{\rm o},\downarrow}) \nonumber \\
    && -  \frac{U}{4}\sum_{\sigma_{1}\sigma_{2}\sigma_{3}\sigma_{4}}
    (\vec{\sigma})_{\sigma_{1}\sigma_{2}} \cdot (\vec{\sigma})_{\sigma_{3}\sigma_{4}}
    d_{{\rm e},\sigma_{1}}^{\dagger} d_{{\rm e},\sigma_{2}} d_{{\rm o},\sigma_{3}}^{\dagger} d_{{\rm o},\sigma_{4}}
    \nonumber \\
    && - \frac{U}{2}(d_{{\rm e},\uparrow}^{\dagger} d_{{\rm e},\downarrow}^{\dagger}
    d_{{\rm o},\downarrow} d_{{\rm o},\uparrow}
    + d_{{\rm o},\uparrow}^{\dagger} d_{{\rm o},\downarrow}^{\dagger} d_{{\rm e},\downarrow} d_{{\rm e},\uparrow}), 
\label{eq:H_d} \\
    H_{{\rm l-d}} & = &
    2 \sum_{k\sigma}
      (
	V \cos\frac{\varphi}{4} d_{{\rm e},\sigma}^{\dagger} \alpha_{{\rm s},k\sigma} 
      + V \sin\frac{\varphi}{4} d_{{\rm o},\sigma}^{\dagger} \alpha_{{\rm a},k\sigma} 
    \nonumber \\
    &&  + h.c. 
       ),
\label{eq:H_l-d_mod}
\end{eqnarray}
where 
$ 
  \alpha_{{\rm s},k\sigma}
  \equiv
  (a_{k\sigma} + b_{k\sigma}) / \sqrt{2}
$
is the symmetric combination of the lead orbits, 
and
$
 \alpha_{{\rm a},k\sigma}
 \equiv
 (a_{k\sigma} - b_{k\sigma}) / \sqrt{2}
$
is the anti-symmetric combination of those,
and 
$ d_{{\rm e},\sigma}
 \equiv
 (d_{1 \sigma} + d_{2 \sigma}) / \sqrt{2}
$
,
$ d_{{\rm o},\sigma}
 \equiv
 {\rm -i} (d_{1 \sigma} - d_{2 \sigma}) / \sqrt{2}
$
are those of the dot orbits,
respectively.
The quantity $\vec{\sigma}$ is the Pauli matrix.
The problem is reduced to solve the two-channel Anderson model
as seen in 
eqs. ({\ref{eq:H_l_mod}}),
({\ref{eq:H_d}}) and
({\ref{eq:H_l-d_mod}}).
Hereafter,
we call the orbit denoted by $d_{{\rm e},\sigma}$ as `even-orbit'
and the channel of it as `even-channel',
and those of $d_{{\rm o},\sigma}$ as `odd-orbit' and `odd-channel'.
We denote them by $l$ as $l={\rm e}$ or $l={\rm o}$.


\subsection{Conductance formula}

In this paper,
we restrict ourselves to the linear response conductance
for the applied bias voltage.
We derive the conductance formula 
without assumptions on electron tunneling processes.
We also derive the conductance formula 
with assumption of the coherent resonant tunneling process.
Calculated results are compared
at very low temperature.


\subsubsection{Conductance formula at finite temperature}

Now, 
we must calculate the conductance 
without using approximations for tunneling processes such as 
the sequential tunneling process picture or the coherent tunneling process picture.
We define an electric current from lead A to lead B as follows; 
\begin{eqnarray}
 I & \equiv & -e \frac{ -\langle \dot{N_{\rm A}} \rangle
  +\langle \dot{N_{\rm B}} \rangle }{2},
  \label{eq:def:current}
\end{eqnarray}
where $-e$ is the charge of an electron,
and 
$\langle \dot{N_{\rm A}} \rangle$ is the expectation value
of the time differentiation of 
$
N_{\rm A}=\sum_{k \sigma} a_{k \sigma}^{\dagger} a_{k \sigma}
$,
the electron number operator in the lead A.
The quantity $\langle \dot{N_{\rm B}} \rangle$ is that of the lead B.
We consider the situation 
applying the voltage $2V$ 
between lead A and lead B.
This situation is given by adding the term
$H' = N_{\rm A} eV -  N_{\rm B} eV$
to our Hamiltonian $H$,
eq. (\ref{eq:totalH}).
Assuming that the voltage difference $2V$ is small,
and using linear response theory derived by Kubo,~\cite{rf:linear}
we obtain the following expression
for the conductance formula (see Appendix);
\begin{eqnarray}
  G
  & \equiv & \frac{I}{2V} 
  \nonumber \\
  & = & \frac{2 e^{2}}{h} 
        \lim_{\omega \rightarrow 0} 
        \frac{P''(\omega)}{\omega},
  \label{eq:G}
\end{eqnarray}
with
\begin{eqnarray}
  P''(\omega) 
  & = & 
  \frac{\pi^{2} \hbar^{2}}{4}
  \frac{1}{Z} \sum_{n,m} 
  \left( e^{- \beta E_{m}} - e^{- \beta E_{n}} \right) 
  \nonumber \\
  && \times 
  \left| \langle n \left| 
  \dot{N_{\rm A}} - \dot{N_{\rm B}} 
  \right| m \rangle \right| ^{2} 
  \nonumber \\
  && \times \delta \left( \omega - ( E_{n} - E_{m} ) \right),
  \label{eq:P''}
\end{eqnarray}
where $Z = \sum_{n} e^{- \beta E_{n}}$
is the partition function,
$\beta=1/T$, 
$T$ is the temperature of the system.

The quantities 
$\langle \dot{N_{\rm A}} \rangle$ and $\langle \dot{N_{\rm B}} \rangle$
can be expressed by localized operators near the dots as shown in Appendix.
So,
this expression is suitable for the numerical calculation by the NRG method,
though it needs the delicate limiting process $\omega \rightarrow 0$.


\subsubsection{Conductance formula at zero temperature 
based on the coherent resonant tunneling process}

Here we derive the conductance formula at zero temperature
for comparison with numerical results from eqs. (\ref{eq:G}) and (\ref{eq:P''}).
The conductance at zero temperature is given by
\begin{eqnarray}
     G_{\rm F} & = &
                      \frac{2\pi e^{2}}{\hbar}
                      \sum_{\sigma,S_{{\rm int}}} \sum_{\sigma^{'},S_{{\rm fin}}}
                      \left| 
                             T_{{\rm fin},{\rm int}}(\varepsilon_{\rm F})
                      \right|^{2} 
                      \nonumber \\
                     && \times \rho^{2} (\varepsilon_{\rm F})
                      W_{S_{{\rm int}}},
                      \label{eq:G-t-mtx}
\end{eqnarray}
where initial and final states of transition are denoted as 
${\rm int}=(k\sigma,S_{{\rm int}})$ and ${\rm fin}=(p\sigma^{'},S_{{\rm fin}})$.
The quantity $k$ denotes the electron in the lead A,
and $p$ denotes that in the lead B.
By $S_{{\rm int}}$ and $S_{{\rm fin}}$
we represent initial and final states on the quantum dots,
and 
$W_{S_{{\rm int}}}$ is the probability of state in $S_{{\rm int}}$.
The quantity $T_{{\rm fin},{\rm int}}$ is the transition matrix,
and $\rho (\varepsilon_{\rm F})$ 
is the density of states on the Fermi energy in the leads.
The transition matrix is written by using Green's function ${\cal G}$ as
$T = V + V {\cal G} V$,
where $V = H_{\rm l-d}$ in this case.

At zero temperature,
the system written in eq. (\ref{eq:totalH})
is expected to be in the local Fermi liquid state.
Electron tunneling processes through the quantum dots 
will be given by only the coherent resonant tunneling process,
i.e.,
$S_{{\rm int}} = S_{{\rm fin}}$.
The conductance $G_{\rm F}$ is written as follows;
\begin{eqnarray}
  G_{\rm F}
  & = &
  \frac{2 e^{2}}{h}
	\left| \Delta_{{\rm e}} {\cal G}_{{\rm e},\sigma}(\varepsilon_{\rm F}+{\rm i}0)
	-      \Delta_{{\rm o}} {\cal G}_{{\rm o},\sigma}(\varepsilon_{\rm F}+{\rm i}0)
	\right|^{2},\nonumber \\
  &&\label{eq:Gif}
\end{eqnarray}
with
\begin{eqnarray}
   \Delta_{{\rm e}} & = & 4 \pi |V|^{2} \rho(\varepsilon_{\rm F}) \cos^{2} \frac{\varphi}{4}, \\
   \Delta_{{\rm o}} & = & 4 \pi |V|^{2} \rho(\varepsilon_{\rm F}) \sin^{2} \frac{\varphi}{4}.
\end{eqnarray}
In the local Fermi liquid state,
Green's function 
${\cal G}_{l,\sigma}(\varepsilon_{\rm F}+ {\rm i} 0)$
in the $l$-th channel $(l={\rm e,o})$ satisfies 
the Friedel sum rule as follows~\cite{rf:Friedel_L,rf:Friedel};
\begin{eqnarray}
 {\cal G}_{l,\sigma}(\varepsilon_{\rm F}+ {\rm i} 0)
  & = &
 \frac{1}
 {
 \left(
  -\Delta_{l} \frac{1}{\tan\delta_{l}} + {\rm i} \Delta_{l}
 \right)
  },
\label{eq:friedel}
\end{eqnarray}
where $\delta_{l}$ is the phase shift for the $l$-th channel.
Using the relation between the phase shift $\delta_{l}$
and the ground-state occupation number 
$\langle n_{0,l} \rangle$ on the $l$-th orbit,
\begin{eqnarray}
  \langle n_{0,l} \rangle & = & 2 \frac{\delta_{l}}{\pi},
\end{eqnarray}
we get the conductance as follows;
\begin{eqnarray}
 G_{\rm F}
  & = &
  \frac{2 e^{2}}{h}
  \sin^{2}
  \left\{
   \frac{\pi}{2}
   (\langle n_{0,{\rm e}} \rangle - \langle n_{0,{\rm o}} \rangle )
   \right\}.
\label{eq:gamma_el}
\end{eqnarray}
Although the expression ({\ref{eq:gamma_el}})
seems to be different from eq. ({\ref{eq:G}}),
it can be derived from eq. ({\ref{eq:G}})
when the system is in the local Fermi liquid state
and $T=0$.

The expression ({\ref{eq:gamma_el}})
is contrast with the single dot case.
In such a case,
the conductance is given by 
$(2 e^2 / h)\sin^{2} (\pi \langle n_{0} \rangle / 2)$,
and it has almost the maximum value $2 e^2 / h$ 
when there is the Kondo resonance peak on the Fermi energy,
i.e., 
$\langle n_{0} \rangle \sim 1$.~\cite{rf:Kondo_dot_Ng,rf:Kondo_dot_JETP,rf:Kondo_dot_Kawa}
On the other hand,
the conductance has very small value in the present case 
as seen from ({\ref{eq:gamma_el}})
if the even and odd components have 
simultaneously the Kondo resonance on the Fermi energy,
i.e., 
$\langle n_{0,{\rm e}} \rangle \sim \langle n_{0,{\rm o}} \rangle \sim 1$.
This is caused by the interference cancellation 
between processes through the even and odd orbit states.


\subsection{Excitation spectra}

We will calculate excitation spectra
other than $P^{''}(\omega)$
to get insights of the electronic states of the dots system.
The single particle excitation spectrum for the 
$l$-th orbit $(l={\rm e,o})$ is defined as follows;
\begin{eqnarray}
 \rho_{l}(\omega)
  & \equiv &
  \frac{1}{Z}
  \sum_{n,m}
  \left(
   e^{-\beta E_{m}} - e^{-\beta E_{n}}
  \right) \nonumber\\
 &&
  \times
  \left(
  \left| \langle n | d_{l}^{\dagger} | m \rangle \right|^{2}
  \delta \left( \omega-(E_{n}-E_{m}) \right) \right.
\nonumber \\
&&  \left. +
  \left| \langle n | d_{l} | m \rangle \right|^{2}
  \delta \left( \omega+(E_{n}-E_{m}) \right)
 \right).
\end{eqnarray}

The magnetic excitation spectrum at $T=0$ is calculated as follows;
\begin{eqnarray}
 \chi_{\rm m,e}^{''}(\omega)
 & \equiv &
 \sum_{n} \sum_{{\rm Gr}}
 \left| \langle n |
  ( S_{{\rm e},z} + S_{{\rm o},z} )
  | {\rm Gr} \rangle \right|^{2}
\nonumber \\
 && \times \delta \left( \omega-(E_{n}-E_{{\rm Gr}}) \right),
\label{eq:chi}
\end{eqnarray}
where 
$S_{l,z} 
= (d_{l,\uparrow}^{\dagger} d_{l,\uparrow} 
 - d_{l,\downarrow}^{\dagger} d_{l,\downarrow})/2$
is the spin operator on the $l$-th orbit and 
${\rm Gr}$ denotes the ground state of the system.
The energy of the peak position will reflect
the characteristic energy of the spin fluctuation.
In this paper,
the energy of the peak position 
of $\chi_{\rm m,e}^{''}(\omega)$ 
is used as the definition 
of the Kondo temperature,
$T_{\rm K}$.
For the four fold degenerate model,
this definition gives rather well approximate value of $T_{\rm K}$
defined by the usual definition based on the susceptibility.~\cite{rf:NRG_sakai}

%

%
\section{Numerical Results}
\label{cal}

We have several parameters for the model.
The band width of the leads, 
$D$, 
is chosen to be energy units,
$D=1$.
The parameters 
$\varepsilon_{\rm d},
U,
\Delta \equiv 4 \pi |V|^{2} \rho(\varepsilon_{\rm F})$ and 
$T$ are given 
in units of $D$.

In this paper,
we fix the Coulomb repulsion energy to be $U = 0.10$
and 
the tunneling intensity $\Delta = 0.03 \pi$.
It is enough to study the only $\varepsilon_{\rm d} \ge -0.5U$ cases 
because the conductance as a function of $\varepsilon_{\rm d}$
is symmetric with respect to $\varepsilon_{\rm d} = -0.5U$.
Then the energy level $\varepsilon_{\rm d}$ 
is varied from $-0.05 = -0.50U$,
corresponding to the electron-hole symmetric case,
to $0.1 = 1.0U$.
The conductance is the periodic function of the magnetic flux $\varphi$
and moreover it is even function of $\varphi$ for the present model,
so we calculate the conductance only in a half period,
$0 \le \varphi \le \pi$.
Temperature $T$ is varied from $5.8 \times 10^{-7}$ to $3.1 \times 10^{-1}$.
(See Table {\ref{table:ed}} and {\ref{table:T}}.)
\begin{table}[hbt]
  \caption
  {
   Parameters 
   $\varepsilon_{\rm d}$,
   $U$,
   $\Delta \equiv 4 \pi |V|^{2} \rho(\varepsilon_{\rm F})$
   in units of the band width $D$.
   }
   \label{table:ed}
   \begin{tabular}{@{\hspace{\tabcolsep}\extracolsep{\fill}}ccc} \hline
	$\varepsilon_{\rm d}$ & $U$    & ${\Delta}/{\pi}$ \\ \hline
	$-0.050 = -0.50U$     &        &                  \\
	$-0.030 = -0.30U$     &        &                  \\
	$-0.025 = -0.25U$     &        &                  \\
	$-0.020 = -0.20U$     & $0.10$ & $0.03$           \\
	$+0.000 = +0.00U$     &        &                  \\
	$+0.050 = +0.50U$     &        &                  \\
	$+0.100 = +1.00U$     &        &                  \\ \hline
   \end{tabular}
\end{table}
\begin{fulltable}[hbt]
   \begin{center}
   \caption
   {
    Temperatures denoted as $T_{i}$ in this paper.    
    These are given in units of the band width $D$.
   }
   \label{table:T}
	\begin{fulltabular}{@{\hspace{\tabcolsep}\extracolsep{\fill}}ccccccc} \hline
	$T_{1}$              & $T_{2}$               & $T_{3}$               & 
	$T_{4}$              & $T_{5}$               & $T_{6}$               &
	$T_{7}$              \\ \hline
	$5.8 \times 10^{-7}$ &  $1.7 \times 10^{-6}$ &  $5.2 \times 10^{-6}$ &
	$1.6 \times 10^{-5}$ &  $4.7 \times 10^{-5}$ &  $1.4 \times 10^{-4}$ &
	$4.2 \times 10^{-4}$ \\ \hline
	\vspace{5mm}
	\hline
	$T_{8}$              & $T_{9}$               & $T_{10}$              & 
	$T_{11}$             & $T_{12}$              & $T_{13}$              &
	                     \\ \hline
	$1.3 \times 10^{-3}$ &  $3.8 \times 10^{-3}$ &  $1.1 \times 10^{-2}$ &
	$3.4 \times 10^{-2}$ &  $1.1 \times 10^{-1}$ &  $3.1 \times 10^{-1}$ &
	                     \\ \hline
   \end{fulltabular}
   \end{center}
\end{fulltable}

Before going to detailed discussions of the numerical results,
we stress the following fact.
The Kondo effects in different channels seem to occur 
as if they are independent 
though the original Hamiltonian includes interaction terms
between electrons in even and odd orbits.
(See eq. (\ref{eq:H_d}).)
As shown later in \S {\ref{sec:cond_finitT}},
the Kondo resonance peaks of the single particle excitation in even-channel 
do not have structure reflecting the energy scale of the odd-channel Kondo effect,
and {\it vice versa}.


\subsection{Conductance at zero temperature $G_{\rm F}(\varphi)$}

\begin{figure}[htb]
  \begin{center}
  \epsfile{file=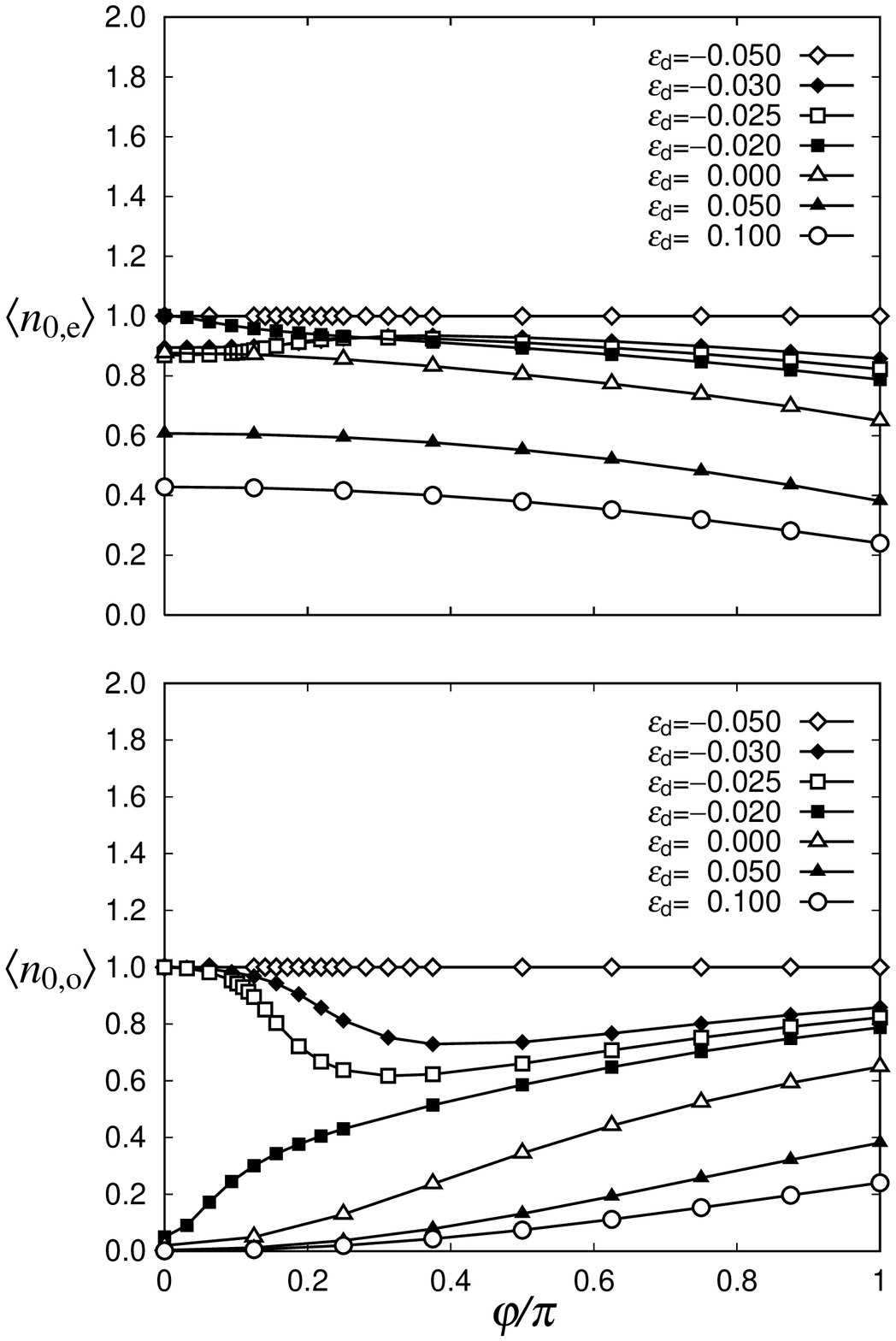,scale=0.55}
  \caption
  {
  The expectation values of the occupation number of the 
  even-orbit,  
  $\langle n_{0,{\rm e}} \rangle$,
  and the odd-orbit,
  $\langle n_{0,{\rm o}} \rangle$,
  in the ground state as a function of 
  $\varphi$ for various cases of $\varepsilon_{\rm d}$.
  The flux $\varphi$ is given in units of $\Phi_{0}=hc/e$.
  The value
  $\langle n_{0,{\rm o}} \rangle$ at $\varphi \sim 0$ decreases rapidly 
  as $\varepsilon_{\rm d}$ increases 
  between $\varepsilon_{\rm d}=-0.025$ to $-0.020$.
  }
  \label{fig:exn}
  \end{center}
\end{figure}

The conductance at zero temperature is written 
by using the difference of the occupation numbers,
$\langle n_{0,{\rm e}} \rangle - \langle n_{0,{\rm o}} \rangle$,
as shown in eq. ({\ref{eq:gamma_el}}).
The expectation value of the occupation number
$\langle n_{0,l} \rangle$ is calculated from the wave function 
obtained by the NRG method,
and is shown in Fig. {\ref{fig:exn}}.
In the electron-hole symmetric case,
$\varepsilon_{\rm d} = -0.050$,
the expectation values
satisfy the relation 
$\langle n_{0,{\rm e}} \rangle = \langle n_{0,{\rm o}} \rangle = 1.0$ for any $\varphi$.
When $\varepsilon_{\rm d}$ moves away from 
$\varepsilon_{\rm d} = -0.050$,
both 
$\langle n_{0,{\rm e}} \rangle$ and $\langle n_{0,{\rm o}} \rangle$
deviate from $1.0$ and they depend on the flux $\varphi$.
We note that the hybridization strength for the odd-orbit,
$\Delta_{{\rm o}}=4 \pi |V|^{2} \rho (\varepsilon_{\rm d}) \sin^{2}(\varphi/4)$,
is very small in the $\varphi \sim 0$ region.
Therefore the small change of the parameters 
in a critical region 
causes drastic change of the
ground state properties of the odd-channel.
Reflecting these,
$\langle n_{0,{\rm o}} \rangle$ at $\varphi \sim 0$
decreases rapidly as $\varepsilon_{\rm d}$ increases
from $-0.025$ to $-0.020$.
At the same time the Kondo resonance peak in the odd-channel
rapidly moves away from the Fermi energy.

We have two regimes for the $\varphi$-dependence of 
$\langle n_{0,{\rm o}} \rangle$.
In the $\varepsilon_{\rm d} \le -0.025$ case,
$\langle n_{0,{\rm o}} \rangle$ has value almost $1.0$ at $\varphi \sim 0$.
It initially decreases with increasing $\varphi$,
and then increases gradually.
In the $\varepsilon_{\rm d} \ge -0.020$ case,
$\langle n_{0,{\rm o}} \rangle$ has very small value at $\varphi \sim 0$.
It increases with increasing $\varphi$. 
In contrast to the behavior of $\langle n_{0,{\rm o}} \rangle$,
the occupation number $\langle n_{0,{\rm e}} \rangle$ shows
mild dependence on $\varepsilon_{\rm d}$
and $\varphi$ because the hybridization strength $\Delta_{{\rm e}}$
is relatively large in 
$0 \le \varphi \le \pi$ region.

\begin{figure}[htb]
  \begin{center}
  \epsfile{file=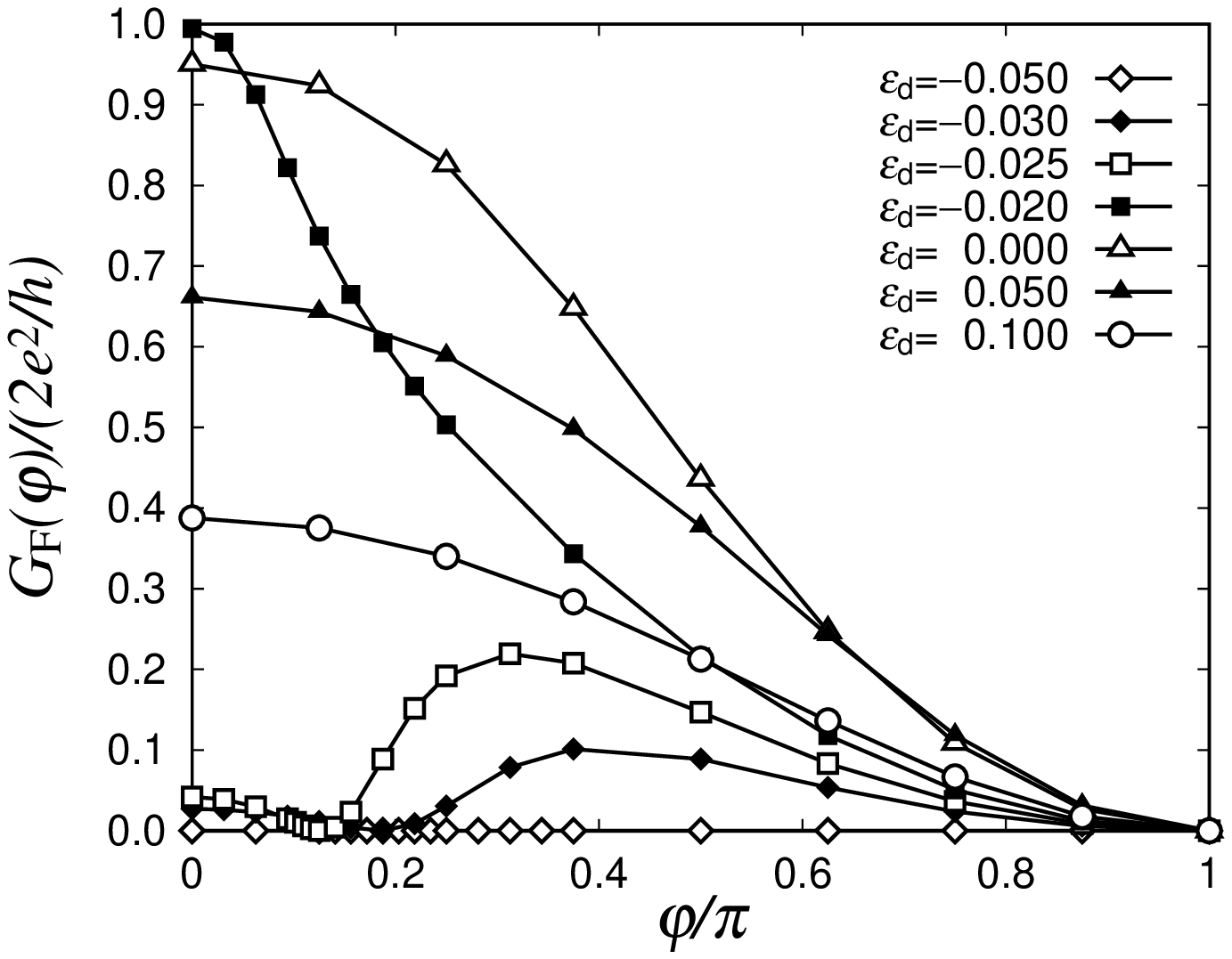,scale=.55}
  \caption
  {
  The conductance at zero temperature, 
  $G_{\rm F}(\varphi)$,
  for various cases of $\varepsilon_{\rm d}$.
  The units of $G_{\rm F}(\varphi)$ is $2 e^{2} / h$.
  In the electron-hole symmetric case ($\varepsilon_{\rm d} = -0.50U = -0.050$),
  AB oscillations vanish completely.
  As $\varepsilon_{\rm d}$ shifts from $\varepsilon_{\rm d} = -0.050$,
  the AB oscillations appear.
  The conductance 
  $G_{\rm F}(\varphi)$ near $\varphi=0$
  increases rapidly 
  when $\varepsilon_{\rm d}$
  increases from $\varepsilon_{\rm d}=-0.025$ to $-0.020$.
  }
  \label{fig:gf}
  \end{center}
\end{figure}

In Fig. {\ref{fig:gf}},
we present
the numerical results
of the conductance at zero temperature $G_{\rm F}(\varphi)$
for various $\varepsilon_{\rm d}$.
The AB oscillations vanish completely in the electron-hole symmetric case,~\cite{rf:n_e=n_o}
$\varepsilon_{\rm d}=-0.050$,
because 
$\langle n_{0,{\rm e}} \rangle = \langle n_{0,{\rm o}} \rangle =1$ as noted in Fig. {\ref{fig:exn}}.
When $\varepsilon_{\rm d}$
moves away from $\varepsilon_{\rm d}=-0.050$,
the AB oscillations appear because
the quantity
$ \langle n_{0,{\rm e}} \rangle - \langle n_{0,{\rm o}} \rangle$
has non zero value and depends on the 
flux $\varphi$.

We note that $G_{\rm F}(\varphi)$ near $\varphi \sim 0$ increases rapidly 
as $\varepsilon_{\rm d}$ increases
between $\varepsilon_{\rm d}=-0.025$ to $-0.020$.
This phenomenon is reflecting the behavior of 
$\langle n_{0,{\rm o}} \rangle$.
In the region of $\langle n_{0,{\rm o}} \rangle \sim 0$,
the Kondo resonance peak of the odd-channel moves away
from the Fermi energy.
Therefore,
the contribution of the resonant tunneling via the Kondo resonance of the odd-channel
becomes very small.
From this we can see that the conductance
$G_{\rm F}(\varphi=0)$ take value near $2 e^{2} / h$
when only the coherent resonant tunneling of the even-channel contributes.

To conclude,
the AB oscillations of the conductance at zero temperature 
show two types of $\varphi$ dependence 
reflecting the change of the ground state properties of the system.
In the $\varepsilon_{\rm d} \le -0.025$ case,
$G_{\rm F}(\varphi)$ is very small at $\varphi \sim 0$.
It decreases initially and soon increases with $\varphi$,
and then it decreases near $\varphi / \pi \sim 1$ after showing maximum.
In other word,
the AB oscillations have higher harmonics components.
On the other hand in $\varepsilon_{\rm d} \ge -0.020$ case,
$G_{\rm F}(\varphi)$ has large value at $\varphi=0$,
and it decreases with $\varphi$.
%


\subsection{Conductance at finite temperature $G(\varphi)$}
\label{sec:cond_finitT}

In this subsection
we show the numerical results of the conductance $G(\varphi)$
at various temperatures
calculated by eqs. ({\ref{eq:G}}) and ({\ref{eq:P''}}).

\begin{figure}[htb]
  \begin{center}
  \epsfile{file=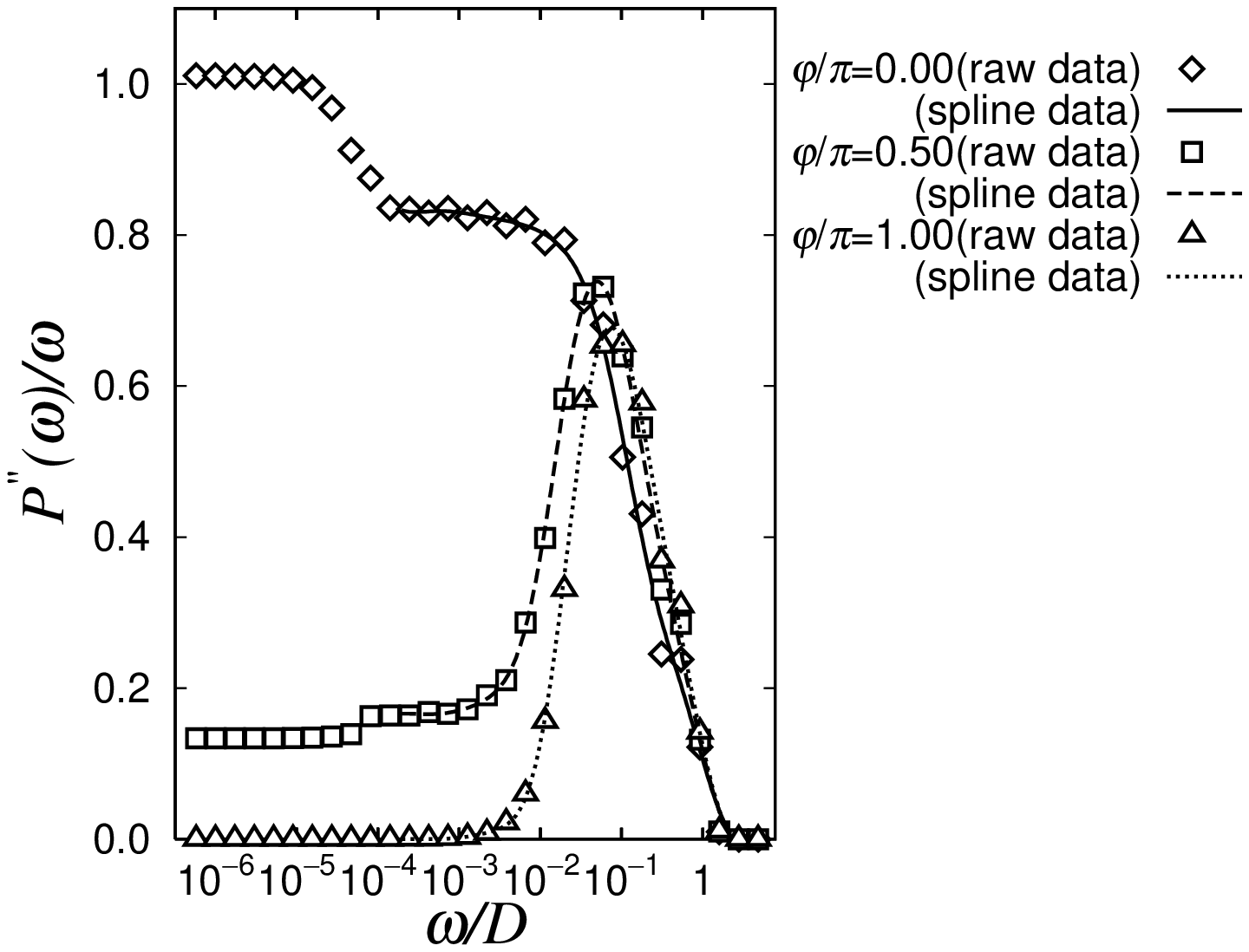,scale=.6}
    \caption
    {
     Examples of the spectrum,
     $P^{''}(\omega) / \omega$,
     calculated 
     by using the NRG method.
     The parameters are 
     $\varepsilon_{\rm d}=-0.025$ and $T=T_{5} (=4.7 \times 10^{-5})$.
     The symbols give the NRG raw data points,
     and lines are given by spline interpolation.
     The edge of the lines give the lowest energy data points 
     for the interpolation,
     $| \omega_{\min} | = 4.0 T_{5}$.
     The data points $\omega / T < \alpha \sim 1$ should be discarded 
     in the NRG calculation,
     see the text.
    }
   \label{fig:spectrum_ori.06}
  \end{center}
\end{figure}

In Fig. {\ref{fig:spectrum_ori.06}},
we show examples of the spectra,
$P^{''}(\omega) / \omega$,
calculated by the NRG method.
The calculated raw data are plotted by symbols,
and the lines give the values 
averaged by the spline interpolation method.
The data points with energies $\omega / T < \alpha \sim 1$
are expected to be erroneous because 
they are calculated from transitions with excitation energy
($T \sim |t_{L}|$)
larger than $\omega$,
where $|t_{L}|$ is the hopping matrix characterizing the 
low energy scale of the NRG calculation.~\cite{rf:NRG_sakai,rf:NRG_shimizu}
In the interpolation,
these low energy data are discarded.
The value at $\omega_{\min}=4.0T$ is substituted for the limiting value of 
$\omega \rightarrow 0$.
As shown from Fig. {\ref{fig:spectrum_ori.06}},
this process does not lead 
the ambiguity of the limiting value of $P^{''}(\omega) / \omega$
for the low temperature cases $T \ll T_{\rm K}$.
But it causes ambiguity for the $T > \alpha^{'} \sim T_{\rm K}$ cases.
The quantitative accuracy 
of the numerical value of the present work should not be
so trusted for higher temperature cases.
However,
it seems not bad for the qualitative discussions.

In the following subsections,
we show two typical cases for the numerical results 
of the conductance at finite temperature.
%


\subsubsection
{
Deep $\varepsilon_{\rm d}$ case
($\varepsilon_{\rm d} = -0.025$)
}\label{sec:ed=-0.025}

\begin{figure}[htb]
  \begin{center}
  \epsfile{file=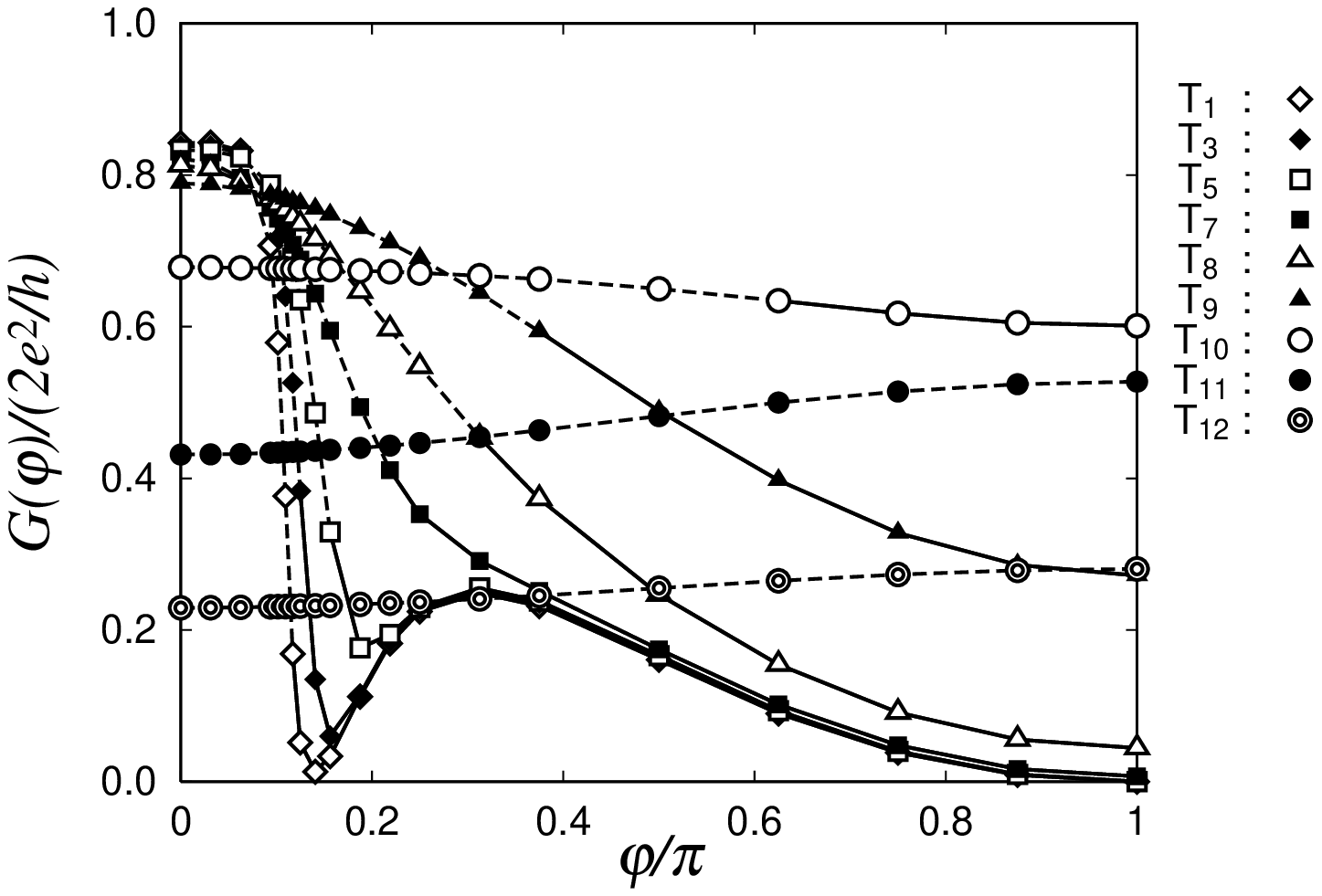,scale=.6}
  \caption
  {
   The conductance $G(\varphi)$
   at $\varepsilon_{\rm d} = -0.025$ for various temperatures.
   The regions given by the broken lines
   are the magnetic state and the regions given by the solid lines
   are the non-magnetic state.
   For the definition,
   see the text.
   The curves of the conductance given by the solid line region 
   for $T_{1} \le T \le T_{7}$ cases coincide to 
   $G_{\rm F}(\varphi)$ of  $\varepsilon_{\rm d} = -0.025$ case in Fig. {\ref{fig:gf}}.
   Tunneling processes
   are expected to be dominated 
   by the coherent resonant tunneling process on solid lines.
   On broken lines, 
   contributions from the process with the spin inelastic scattering will be not so small.
   The conductance rapidly changes at $\varphi / \pi \sim 0.10$ 
   in the temperature cases $T_{1} \le T \le T_{7}$.
   For the origin of this phenomena, 
   see also the text.
  }
  \label{fig:cond-.0250}
  \end{center}
\end{figure}

Numerical results of 
the conductance $G(\varphi)$
at $\varepsilon_{\rm d} = -0.025$
for various temperatures
are shown in Fig. {\ref{fig:cond-.0250}}.

First,
we discuss the results
in the temperature range $T_{1} \le T \le T_{7}$.
The conductance $G(\varphi)$ changes rapidly
as $\varphi$ changes between $0.1 < \varphi / \pi  < 0.3$.
In addition,
it has value near $2 e^{2} / h$
in $\varphi / \pi < 0.1$ region
contrasted to the result of $G_{\rm F}(\varphi)$
for $\varepsilon_{\rm d}=-0.025$ in Fig. {\ref{fig:gf}}.
In the region of $\varphi / \pi > 0.3$,
the curves of the conductance $G(\varphi)$
overlap to each other,
and also coincide with the result from $G_{\rm F}(\varphi)$.

To get insights the origin of this behavior of the conductance,
we investigate the electronic state of the system.
The one-particle excitation spectra  
$\pi \Delta_{{\rm e}} \rho_{{\rm e}} (\omega)$ and
$\pi \Delta_{{\rm o}} \rho_{{\rm o}} (\omega)$,
and also the current spectra $P^{''}(\omega)$
for various flux cases at $T=T_{1}$
are shown in Fig. {\ref{fig:spectrum.02}}.
\begin{fullfigure}[htb]
  \begin{center}
  \epsfile{file=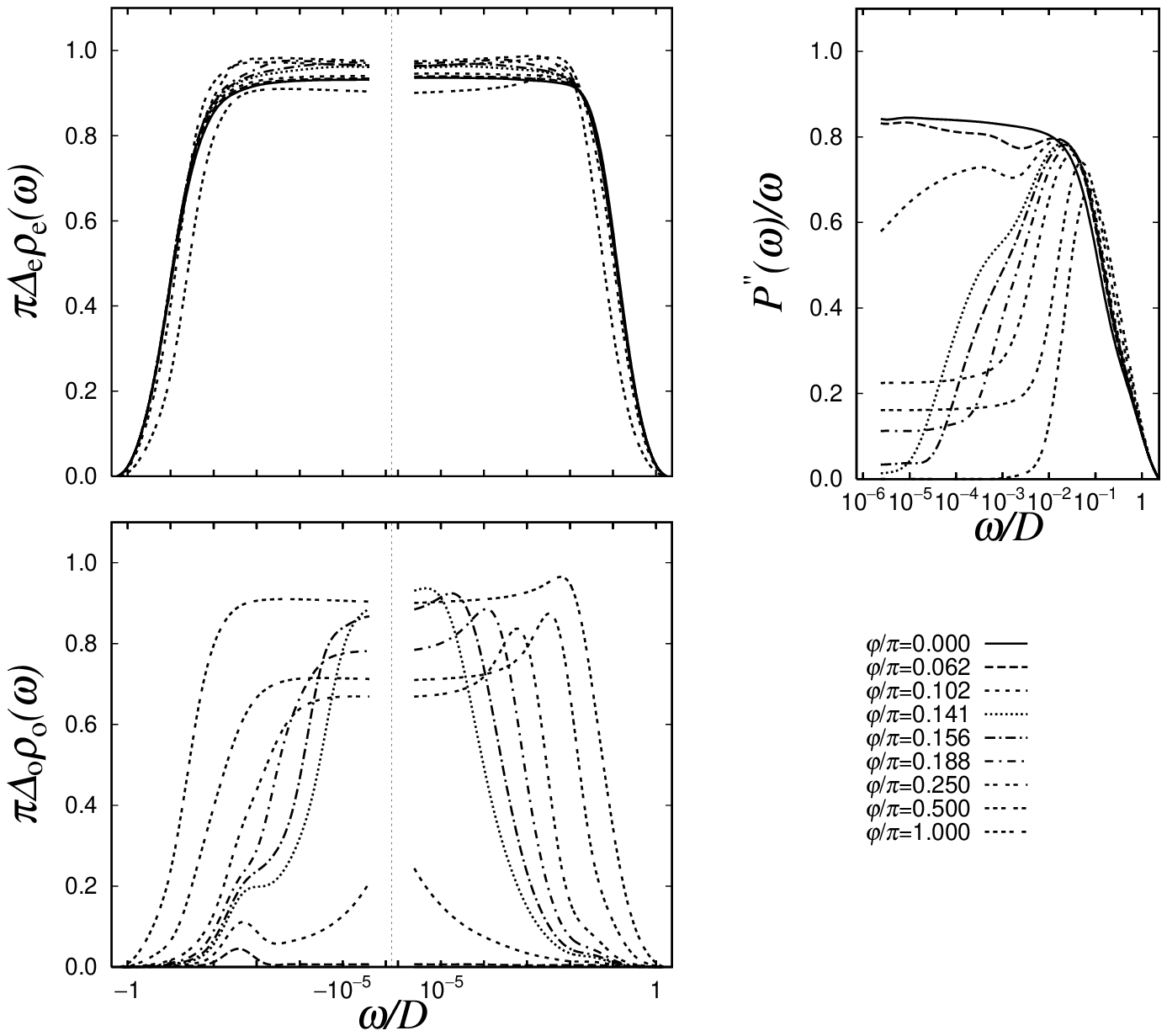,scale=.85}
  \caption
  {
   Excitation spectra 
   $\pi \Delta_{{\rm e}} \rho_{{\rm e}} (\omega)$,
   $\pi \Delta_{{\rm o}} \rho_{{\rm o}} (\omega)$,
   and 
   $P^{''}(\omega) / \omega$ 
   at $T=T_{1}$ and
   $\varepsilon_{\rm d}=-0.025$.
   The Kondo resonance peaks on Fermi energy 
   always exist in $\rho_{{\rm e}} (\omega)$.
   On the other hand, 
   There are no Kondo resonance peaks 
   in $\rho_{{\rm o}} (\omega)$ 
   when $\varphi / \pi < 0.1$.
   The Kondo resonance peaks in $\rho_{{\rm o}} (\omega)$ 
   gradually grow up as flux increases from $\varphi / \pi \sim 0.1$.
   The peaks appear completely in the $\varphi / \pi > 0.15$ region.
  }
  \label{fig:spectrum.02}
  \end{center}
\end{fullfigure}
The same quantities at $T=T_{3}$
are shown in Fig. {\ref{fig:spectrum.04}}.
\begin{fullfigure}[htb]
  \begin{center}
  \epsfile{file=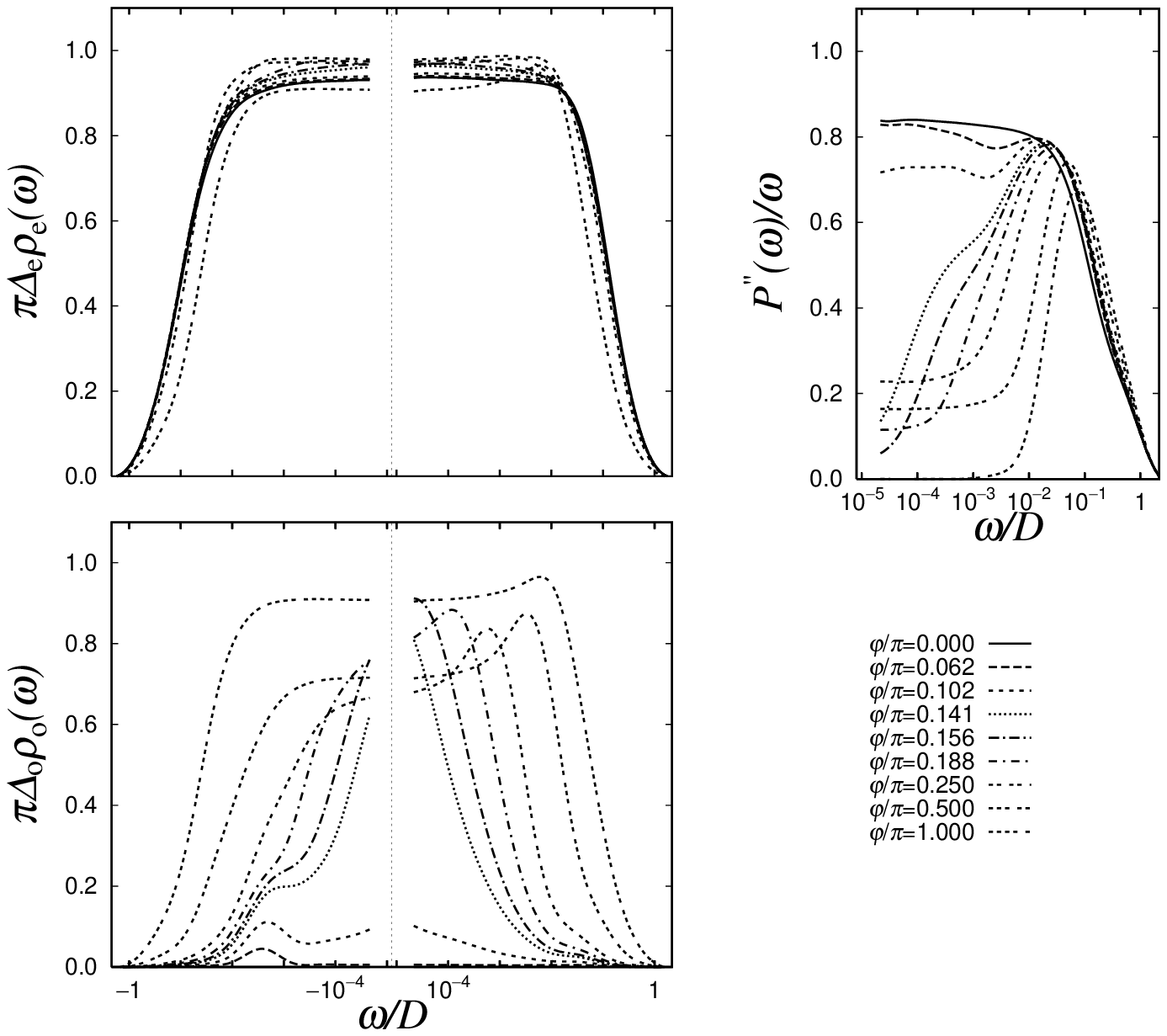,scale=.85}
  \caption
  {
   Excitation spectra 
   $\pi \Delta_{{\rm e}} \rho_{{\rm e}} (\omega)$,
   $\pi \Delta_{{\rm o}} \rho_{{\rm o}} (\omega)$,
   and 
   $P^{''}(\omega) / \omega$ 
   at $T=T_{3}$ and
   $\varepsilon_{\rm d}=-0.025$.
   The Kondo resonance peaks in $\rho_{{\rm o}} (\omega)$ 
   appear in the  $\varphi / \pi > 0.25$ region.
   The region of $\varphi$ that the odd-channel being magnetic state
   becomes wider when the temperature rises
   as seen from comparison with Fig. {\ref{fig:spectrum.02}}.   
  }
  \label{fig:spectrum.04}
  \end{center}
\end{fullfigure}
At $T=T_{1}$,
we have the Kondo resonance peaks 
in the even-channel for all flux $\varphi$ cases.
(Notice that the abscissa have a logarithmic scale.)
On the other hand in $\varphi / \pi < 0.1$ cases for the odd-channel,
the widths of the Kondo resonance peaks
are small compared to the lowest energy calculated 
in the Fig. {\ref{fig:spectrum.02}}.
The width gradually increases as 
flux $\varphi$ increases from $\varphi / \pi  \sim 0.10$,
and the complete Kondo resonance peak
appears in the $\varphi / \pi > 0.15$ cases.

The evolution of the Kondo resonance peak 
in a certain channel means that 
the electronic state in the channel 
is in the spin singlet state.
In this case,
the excitation properties will be described as the Fermi liquid state.
Therefore,
the electron tunneling processes through that channel
will be given by only the coherent resonant tunneling process.

The even-channel state is always in the Fermi liquid state at $T=T_{1}$
for whole range of $\varphi$.
The electronic state of the odd-channel falls into the Fermi liquid state
for $\varphi / \pi > 0.15$ at $T=T_{1}$.
On the other hand for the cases $\varphi / \pi < 0.1$, 
the spin state of the odd-channel at $T=T_{1}$ 
is expected to be magnetic.
Fig. {\ref{fig:spectrum.04}}
shows the spectra at $T=T_{3}$.
The intensity of the Kondo resonance of odd-channel 
with $\varphi / \pi = 0.141$ decreases remarkably
from the lines in Fig. {\ref{fig:spectrum.02}}.
At the same time the conductance increases.

\begin{figure}[htb]
  \begin{center}
  \epsfile{file=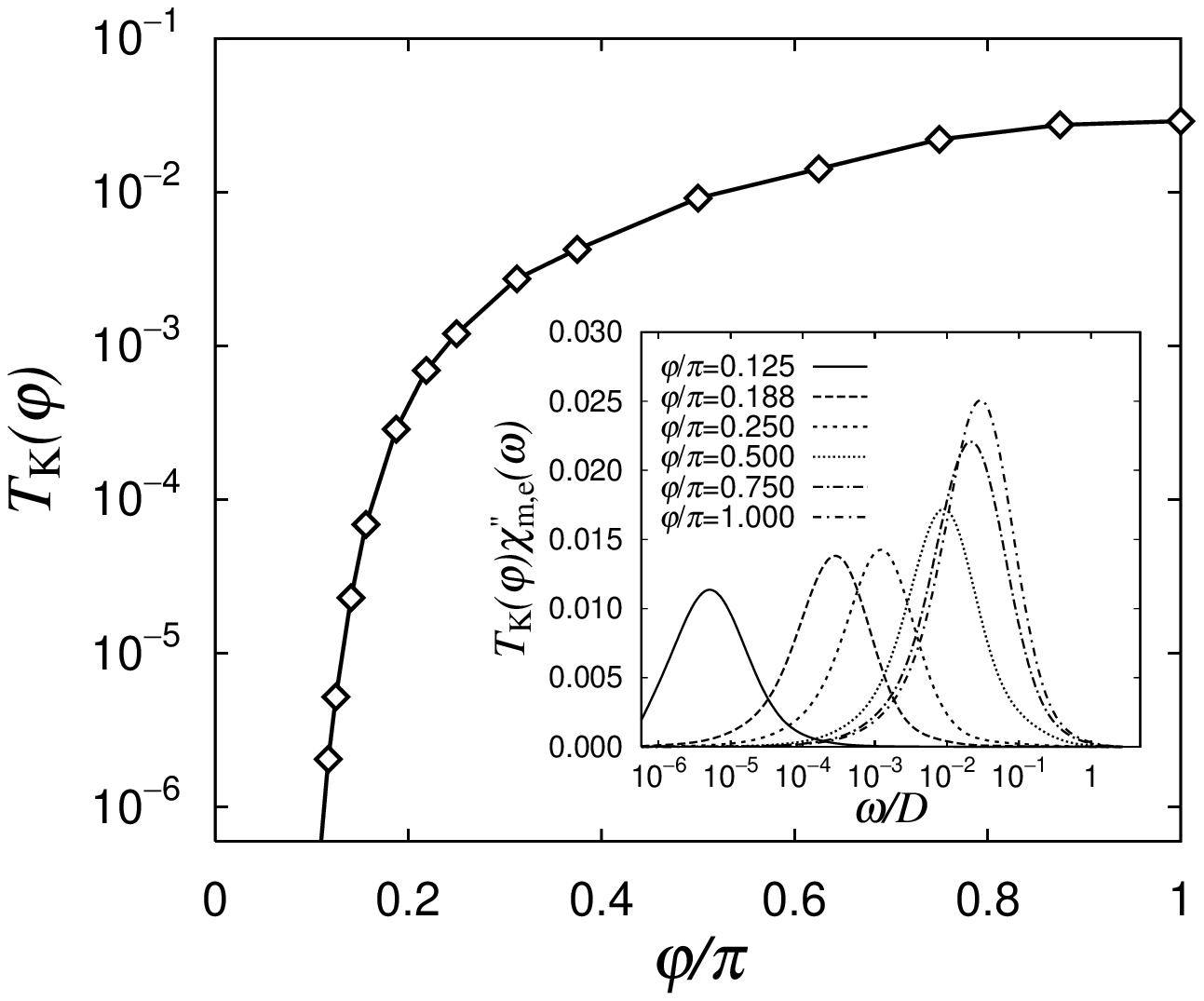,scale=.60}
  \caption
  {
   The Kondo temperature $T_{\rm K}(\varphi)$ 
   as a function of the magnetic flux $\varphi$
   at $\varepsilon_{\rm d}=-0.025$.
   Inset data are the magnetic excitation spectra 
   at zero temperature.
   The energies $\omega$ of the peak positions of the spectra 
   are used for the definition of $T_{\rm K}(\varphi)$.
   $T_{\rm K}(\varphi \sim 0)$ is extremely low 
   because the hybridization intensity of the odd-channel,
   $\Delta_{{\rm o}}$,
   is proportional to $\sin^{2}(\varphi / 4)$
   and is very small for $\varphi \sim 0$.
   $T_{\rm K}(\varphi)$ increases as $\varphi$ increases 
   from $\varphi=0$
   to $\pi$.
  }
  \label{fig:T_Klog-.0250}
  \end{center}
\end{figure}

It may be advisable to consider 
the Kondo temperature,
$T_{\rm K}$,
for given $\varphi$ case.
Now,
we define $T_{\rm K}(\varphi)$ as an energy of 
the peak position of the magnetic excitation spectrum
at zero temperature $\chi_{\rm m,e}^{''}(\omega)$ given by eq. (\ref{eq:chi}).
The result is shown in 
Fig. {\ref{fig:T_Klog-.0250}}.
In the figure we also show the spectrum $\chi_{\rm m,e}^{''}(\omega)$
for several values of $\varphi$.
The energy of the peak position is expected to reflect 
the characteristic energy of the spin fluctuation on the odd-channel.
For $\varphi / \pi =0.125$,
we have peak at $\omega/D = 5.2 \times 10^{-6}$ 
and the energy of the peak position increases rapidly
as $\varphi$ increases.
For the cases $\varphi / \pi < 0.125$,
the peak position shift to very low energy,
and thus we can not show it in the frame of the figure.

The judgments of the boarder
between the magnetic state (broken lines) 
to non-magnetic state (solid lines) in Fig. {\ref{fig:cond-.0250}}
were given by using $T_{\rm K}(\varphi)$ in Fig. {\ref{fig:T_Klog-.0250}}.
Tunneling processes are expected to 
occur by the coherent resonant tunneling process in the solid line region.
On the broken lines,
the contribution from the processes with the inelastic spin excitation
will be not small.
The conductance show relatively large difference from $G_{\rm F}(\varphi)$
in this region.

When the temperature $T$ rises beyond $T_{7}$,
the conductance near $\varphi / \pi \sim 1$ gradually increases and 
the AB oscillations become relatively smaller.
When the temperature increases further through $T_{10}=1.1 \times 10^{-2}$, 
the conductance near $\varphi / \pi \sim 0$
gradually decreases and the AB oscillations become very small.~\cite{rf:-cos}
The spin states of both even and odd channels gradually change 
from non-magnetic to magnetic state for $\varphi / \pi \sim 1$
as temperature rises up.

\begin{figure}[htb]
  \begin{center}
  \epsfile{file=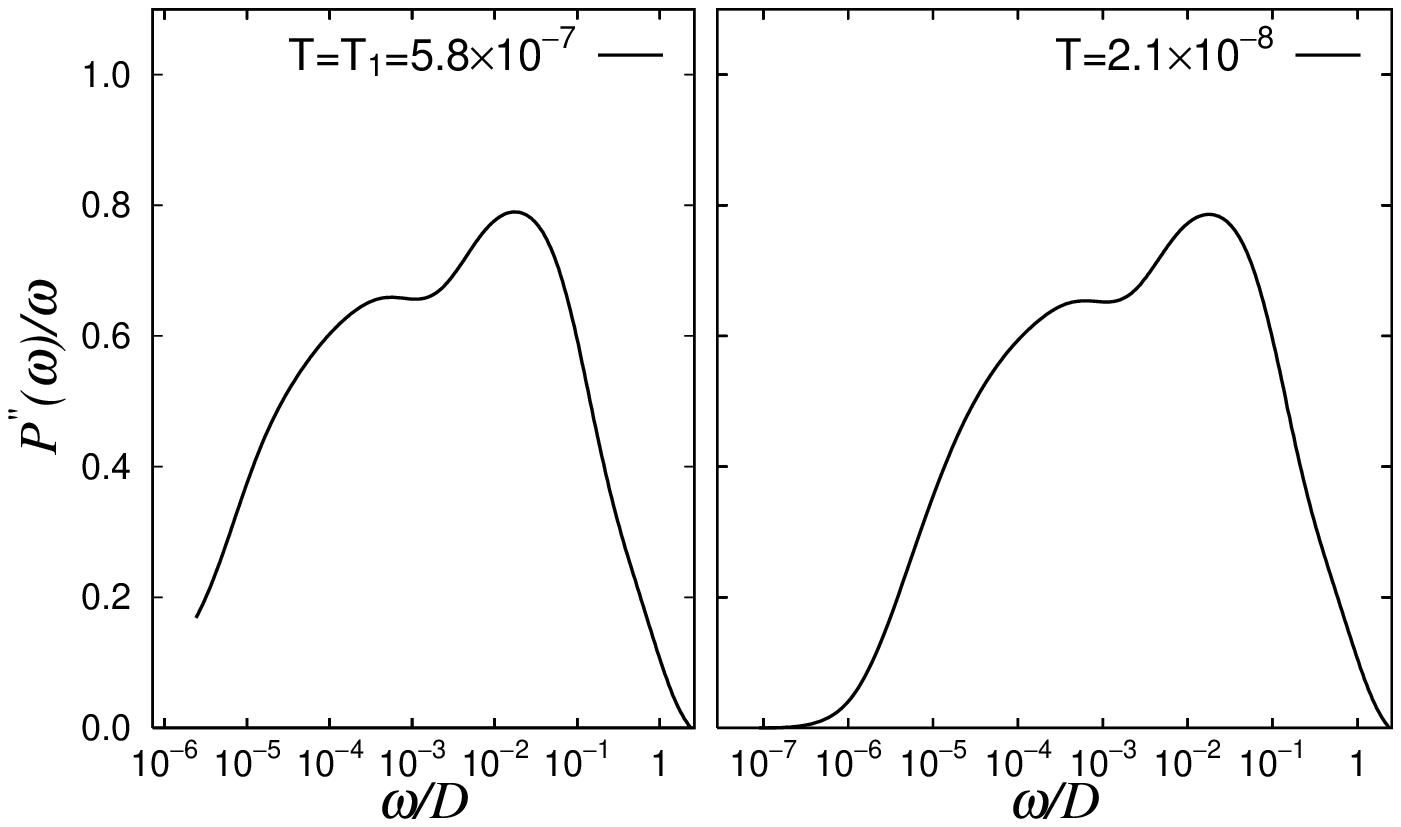,scale=.60}
  \caption
  {
   The excitation spectra $P^{''}(\omega) / \omega$
   at the temperature 
   $T = T_{1} = 5.8 \times 10^{-7}$ (left figure)
   and 
   $T = 2.1 \times 10^{-8}$ (right figure)
   at $\varphi / \pi = 0.117$,
   $\varepsilon_{\rm d} = -0.025$.
   The spectrum $P^{''}(\omega) / \omega$ at $T = 2.1 \times 10^{-8}$
   gives consistent result with 
   $G_{\rm F}(\varphi / \pi = 0.117)=2.4 \times 10^{-3}$ at  
   $\varepsilon_{\rm d} = -0.025$.
  }
  \label{fig:current.cmp}
  \end{center}
\end{figure}

We can conclude that
the behavior of the conductance 
strongly reflects
the spin state in the quantum dots.
At this place we comments on the difference of $G(\varphi)$ 
at very low temperature $T=T_{1}$ from the curve of $G_{\rm F}(\varphi)$
at exactly zero temperature.
The conductance $G_{\rm F}(\varphi = 0)$ is less than $0.1 \times 2 e^{2} / h$,
but $G(\varphi=0)$ is almost $2 e^{2} / h$ even at very low temperature $T=T_{1}$.
In our calculation,
it is very difficult to get the results which agree with $G_{\rm F}(\varphi)$
in the region $\varphi / \pi < 0.15$,
because we must continue the numerical computation to extremely low energy region.
We have checked that calculations up to low energy region 
$\omega \sim 10^{-7}$,
$T\sim 10^{-8}$
gives consistent result with $G_{\rm F}(\varphi)$ for $\varphi / \pi = 0.117$.
(See Fig. {\ref{fig:current.cmp}}).
But it is impossible 
to continue such calculation to $\varphi / \pi < 0.117$ cases.
In the $\varphi / \pi < 0.15$ cases at $T=T_{1}$,
the contribution from the coherent resonant tunneling process of 
the odd-channel will be small 
because $T_{\rm K}(\varphi)$ is less than $T_{1}$.
In such cases,
the conductance will have the value near $2 e^{2} / h$
as seen from Fig. {\ref{fig:gf}}.
We think that abrupt increase of $G(\varphi \sim 0)$ at $T=T_{1}$
from $G_{\rm F}(\varphi \sim 0)$ really reflects the crossover of states 
due to the increase of the temperature.

Here,
we note the meaning of the spectrum $P^{''}(\omega) / \omega$.
This quantity is related to the fluctuation of the bias voltage.
It shows peak structure at energy comparable to the width of the 
Kondo resonance at the Fermi energy.
It shows two peaks structure when $\varphi$ is small, 
as clearly seen in Fig. {\ref{fig:current.cmp}}.
These peaks seem to relate to the Kondo resonance in each channel.


\subsubsection
{
Shallow $\varepsilon_{\rm d}$ case
($\varepsilon_{\rm d} = 0.000$)
}\label{sec:ed=0.000}

\begin{figure}[htb]
  \begin{center}
  \epsfile{file=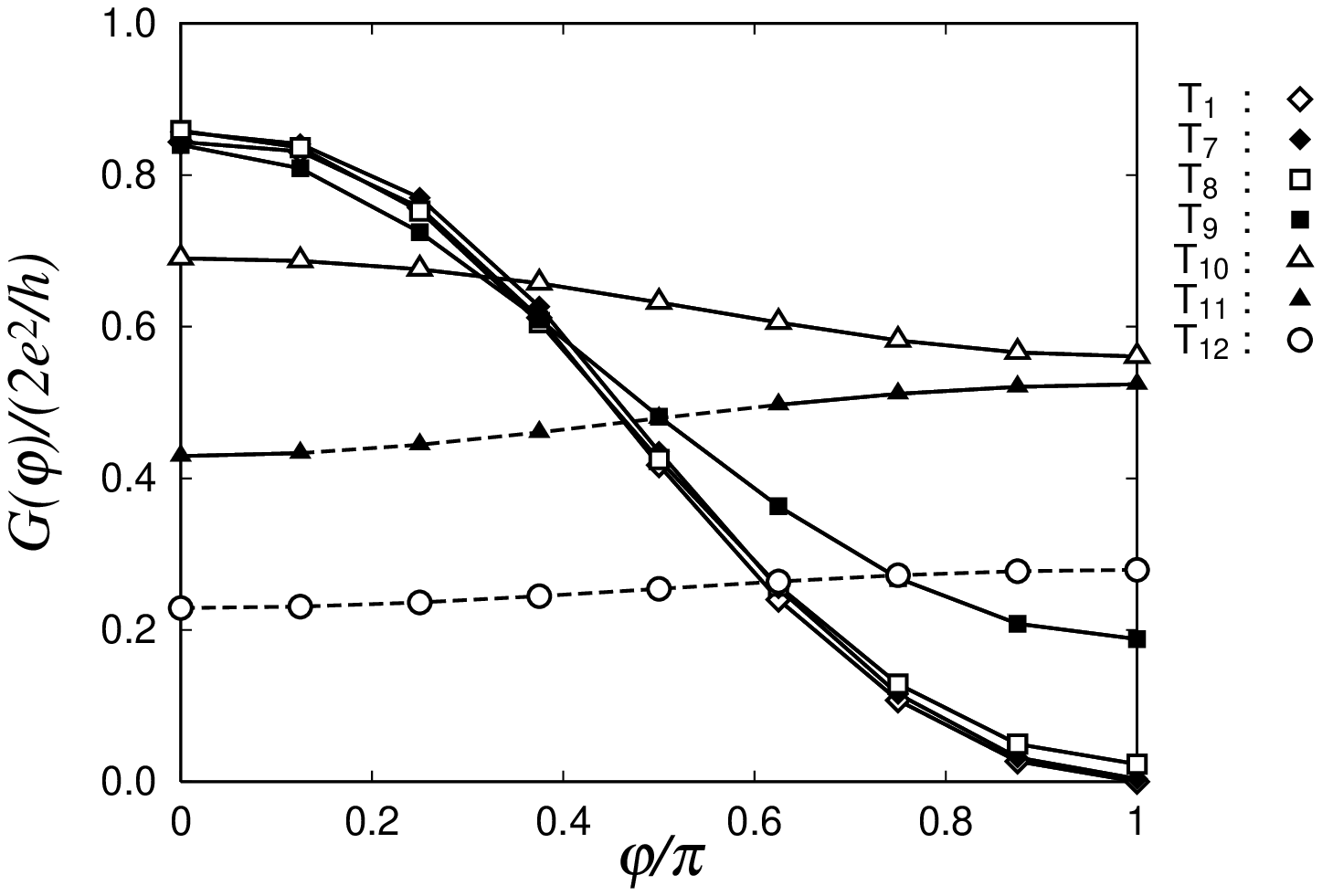,scale=.60}
  \caption
  {
   The conductance $G(\varphi)$ for various temperatures
   at $\varepsilon_{\rm d}=0.000$.
   AB oscillations become smaller 
   as temperature rises up than $T \sim T_{9}$.
   The phenomenon that $G(\varphi)$ rapidly changes with flux
   at low temperatures can not be seen in this case.
   }
  \label{fig:cond-.0000}
  \end{center}
\end{figure}

\begin{figure}[htb]
  \begin{center}
  \epsfile{file=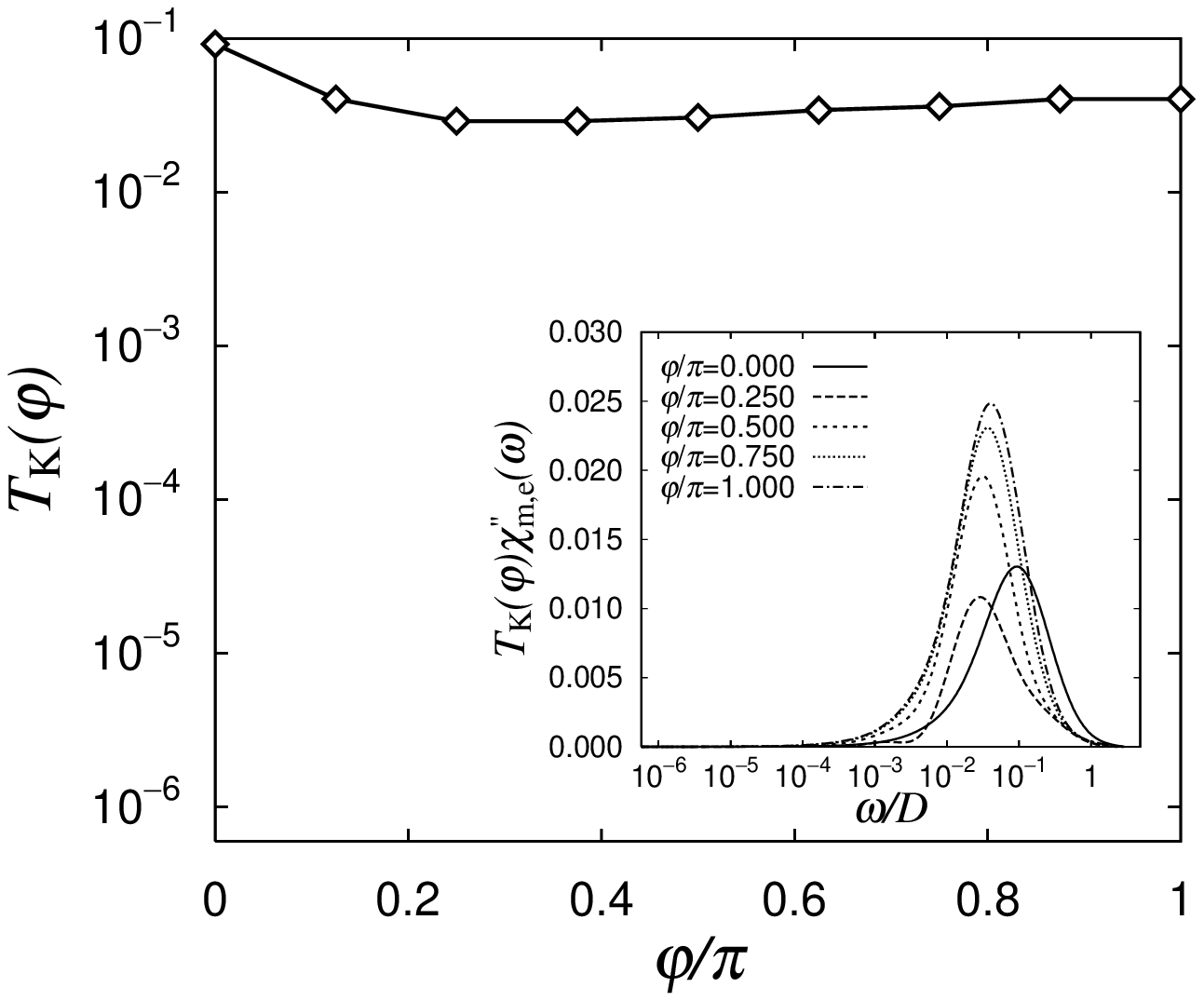,scale=.60}
  \caption
  {
   The Kondo temperature $T_{\rm K}(\varphi)$ 
   as a function of the magnetic flux $\varphi$
   at $\varepsilon_{\rm d}=0.000$ case.
   Inset data are the magnetic excitation spectra 
   at zero temperature.
   The energies $\omega$ of the peak positions of the spectra 
   are used for the definition of $T_{\rm K}(\varphi)$.
   The variation of $T_{\rm K}(\varphi)$ 
   for the change of $\varphi$
   is smaller when compared with result for $\varepsilon_{\rm d}=-0.025$ case.
  }
  \label{fig:T_Klog-.0000}
  \end{center}
\end{figure}

Numerical results of the conductance $G(\varphi)$
at $\varepsilon_{\rm d} = 0.000$ for various $T$
are shown in Fig. {\ref{fig:cond-.0000}}.
The conductance $G(\varphi)$ does not show rapid changes in this case,
contrasted to the results shown in Fig. {\ref{fig:cond-.0250}}.
We show the Kondo temperature $T_{\rm K}(\varphi)$ 
in the $\varepsilon_{\rm d} = 0.000$ case
in Fig. {\ref{fig:T_Klog-.0000}}.
It is much higher than 
that of $\varepsilon_{\rm d} = -0.025$ case in Fig. {\ref{fig:T_Klog-.0250}}
at $\varphi \sim 0$.
In the $\langle n_{0,{\rm o}} \rangle \sim 0.0$ case,
spin freedom on the odd-orbit will disappear
and so
the spin excitation mainly comes from the electrons in the even-orbit.
Then the Kondo temperature $T_{\rm K}(\varphi)$
is larger in this case.
Even when the occupation number on odd-orbit 
increases due to the increase of $\varphi$,
the characteristic magnetic excitation energy 
is not so small because the electrons in both channels
are in the mixed valence regime.
So the variation of 
$T_{\rm K}(\varphi)$ 
for change of $\varphi$
is much smaller than that in the $\varepsilon_{\rm d} = -0.025$ case.
Therefore
the conductance in Fig. {\ref{fig:cond-.0000}}
does not show the rapid change
caused by the crossover of the spin state.
In the temperature range $T<T_{9}$,
the conductance shows simple $\cos \varphi$
like behavior contrasted to the complicated 
$\varphi$ dependence of $\varepsilon_{\rm d}=-0.025$ case.
When the temperature $T$ rises up than $T_{10}$,
the AB oscillations become smaller 
because the temperature becomes comparable to $T_{\rm K}(\varphi)$.~\cite{rf:-cos}

%

%
\section{Summary and Discussion}
\label{sum}

In this paper,
we have investigated the temperature and the magnetic flux dependences 
of the conductance $G$ 
for the AB circuit including two quantum dots.
This quantity is expected to reflect 
the coherency of the electronic state,
and is closely related to the 
`Kondo effect in quantum dots'.
The model Hamiltonian had the form of the two-channel Anderson model
with even and odd orbits.
The conductance was calculated by using the NRG method
based on the formula without assumptions on tunneling processes.

There were interaction terms between electrons in even and odd orbits,
but the Kondo effects in the even-channel and the odd-channel
occurred as if they were independent.
When the dots system were in the Kondo regime near $\varphi \sim 0$,
i.e., the occupation number in the odd-orbit satisfies 
$\langle n_{0,{\rm o}} \rangle \sim 1.0$,
the crossover of the spin state in the odd-channel 
occurred at a critical value of $\varphi$ at low temperature.
This is because the hybridization strength of 
odd-channel increases from the small value near $\varphi \sim 0$
to large value with increasing $\varphi$.
AB oscillations had strong higher harmonics components
reflecting the crossover of the spin state of the system.
This result was contrasted to the single $\cos \varphi$ dependence derived
by Akera.~\cite{rf:AB_dot_Akera}
When the temperature rose up very high,
the amplitude of the AB oscillations became smaller as expected.
When the dots system were in the mixed valence regime near $\varphi \sim 0$,
i.e., 
$\langle n_{0,{\rm o}} \rangle \sim 0.0$,
the drastic change of the conductance as a function of $\varphi$ did not occur.

At this place,
we remark limitation 
of the assumptions used to derive the simplified
expressions,
eqs. ({\ref{eq:H_l_mod}}),
({\ref{eq:H_d}}) and ({\ref{eq:H_l-d_mod}}).
The even and odd channels were separated out,
and the hybridization of odd-channel is proportional to $\sin (\varphi/4)$.
This led the drastic change of the electronic state 
of the odd-channel near $\varphi \sim 0$.
However,
if the dot 1 and dot 2 are asymmetric under the interchange of them,
electrons in both channels mix with each other.
Therefore,
the $\varphi$-dependence near $\varphi \sim 0$ will be blurred
in the asymmetric case.
The experimental work of Yacoby {\it et al.} has been done
in this situation.~\cite{rf:AB_dot_Yacoby,rf:AB_dot_Yacoby2}
Extension of our calculation to the asymmetric cases will be given in near future.
We stress that the present work treated the case that most drastic changes of the
tunneling conductance are expected.

Anyway,
the conductance will show sharp  
$\varphi$-dependence reflecting the crossover of the spin state of dots,
when the Kondo effect is caused by 
the suitable variation of the dot potential energy.

%

%
\section*{Acknowledgments}

The authors would like to thank H. Akera for valuable discussions and information,
and R. Takayama for helpful advice in numerical computation.
The numerical computation was performed 
at the Computer Center of Institute for Molecular Science,
the Computer Center of Tohoku University
and the Supercomputer Center of Institute for Solid State.
%

%

%
\appendix
\section
{Formulation of the conductance in the linear response theory}

We define an electric current as follows;
\begin{eqnarray}
     I & \equiv & -e \frac{ -\langle \dot{N_{{\rm A}}} \rangle
     +\langle \dot{N_{{\rm B}}} \rangle }{2},
 \label{app:eq:current}
\end{eqnarray}
where $-e$ is the charge of an electron.
The quantity 
$
\dot{N_{{\rm A}}}
\equiv
\frac{{\rm i}}{\hbar} [ H , N_{{\rm A}} ]
$
is the time derivative of the electron number operator in the lead A,
$
N_{{\rm A}} 
= 
\sum_{k \sigma}a_{k \sigma}^{\dagger} a_{k \sigma}
$,
and $\dot{N_{{\rm B}}}$ is that in the lead B.

We consider the current 
when the external perturbation term 
$H^{'} = -N_{{\rm A}}V_{\rm A} -N_{{\rm B}}V_{\rm B}$
is added to $H$
which is given by eq. ({\ref{eq:totalH}}).
The expectation values,
$\langle \dot{N_{{\rm A}}} \rangle$ and 
$\langle \dot{N_{{\rm B}}} \rangle$
are given within the linear response theory,~\cite{rf:linear}
\begin{eqnarray}
    \langle \dot{N_{{\rm A}}} \rangle
    & = &
    \sigma_{\rm AA} V_{\rm A} + \sigma_{\rm AB} V_{\rm B}, \\
    \langle \dot{N_{{\rm B}}} \rangle
    & = &
    \sigma_{\rm BA} V_{\rm A} + \sigma_{\rm BB} V_{\rm B},
\end{eqnarray}
with 
\begin{eqnarray}
    \sigma_{\mu \nu} 
    & = &
    \lim_{\omega \rightarrow 0} \sigma_{\mu \nu}(\omega) \\
    & \equiv &
    \lim_{\omega \rightarrow 0}
    \frac{{\rm i}}{\hbar}
    \left\{ 
    \int_{0}^{\infty} dt^{'} 
    e^{-\delta t^{'} +{\rm i} \omega t^{'}}
    \right.     \nonumber \\
    &&
    \left.    
    \times
    {\rm Tr} 
      \left(
           e^{-{\rm i}\frac{H}{\hbar}t^{'}}
           [N_{\nu},\rho_{\rm eq}]
           e^{{\rm i}\frac{H}{\hbar}t^{'}}
           \dot{N_{\mu}}
      \right)
    \right\},
\end{eqnarray}
where $\delta = 0+$.
Using properties of the trace operator
and 
integration by parts,
we get the following expression;
\begin{eqnarray}
    \sigma_{\mu \nu}(\omega) 
    & = &
    \frac{1}{{\rm i} \omega } \left( K_{\mu \nu}(\omega) - K_{\mu \nu}(0) \right),
\end{eqnarray}
with 
\begin{eqnarray}
    K_{\mu \nu}(\omega) & \equiv & - \frac{{\rm i}}{\hbar} \int_{0}^{\infty}
    e^{-\delta t + {\rm i} \omega t}
    \langle [\dot{N}_{\nu} ,\dot{N}_{\mu}(t)] \rangle dt.
\end{eqnarray}
The quantity 
$K_{\mu \nu}(\omega)$ can be expressed by the following form;
\begin{eqnarray}
    K_{\mu \nu}(\omega)
    & = &
    K^{'}_{\mu \nu}(\omega) + {\rm i} K^{''}_{\mu \nu}(\omega),
\end{eqnarray}
where $K^{'}_{\mu \nu}(\omega)$ and $K^{''}_{\mu \nu}(\omega)$ are given
as follows
by using the eigenstates of $H$
(i.e., $H | m \rangle = E_{m} | m \rangle $),
\begin{eqnarray}
    K_{\mu \nu}^{'}(\omega) 
    & = &
    \frac{1}{Z}
    \sum_{n,m}
    \left(
    e^{-\beta E_{n}} - e^{-\beta E_{m}}
    \right)  \nonumber \\
    &&
    \times
    \frac{E_{m}-E_{n}+\hbar\omega}
    {(E_{m}-E_{n}+\hbar\omega)^{2} + (\hbar\delta)^{2}} \nonumber \\
    && \times
    \langle n | \dot{N}_{\nu} | m \rangle  \langle m | \dot{N}_{\mu} | n \rangle \\
    & = &
    K_{\mu \nu}^{'}(-\omega), \\
    K_{\mu \nu}^{''}(\omega) & = &
    \frac{1}{Z}
    \sum_{n,m}
    \left(
    e^{-\beta E_{n}} - e^{-\beta E_{m}}
    \right)  \nonumber \\
    &&
    \times
    \frac{-\hbar\delta}
    {(E_{m}-E_{n}+\hbar\omega)^{2} + (\hbar\delta)^{2}} \nonumber \\
    && \times
    \langle n | \dot{N}_{\nu} | m \rangle  \langle m | \dot{N}_{\mu} | n \rangle
    \label{ImP} \\
    & = &
    -K_{\mu \nu}^{''}(-\omega).
\end{eqnarray}
The quantity $\sigma_{\mu \nu}$ is given by
\begin{eqnarray}
    \sigma_{\mu \nu}
    & = &
    \lim_{\omega \rightarrow 0} \frac{K_{\mu \nu}^{''}(\omega)}{\omega}.
\end{eqnarray}

We give the value
$-V_{\rm A} = -eV$
and
$-V_{\rm B} = eV$
for external field,
and obtain the electric current as follows;
\begin{eqnarray}
    I & = & \frac{e^{2}}{4}
    \left(
    \sigma_{\rm AA} + \sigma_{\rm BB} - \sigma_{\rm AB} - \sigma_{\rm BA}
    \right) 
    \cdot 2V.
\end{eqnarray}
The conductance $G$ is given by
\begin{eqnarray}
    G
    & \equiv &
    \frac{I}{2V} \\
    & = &
    \frac{2 e^{2}}{h}
    \lim_{\omega \rightarrow 0} \frac{P^{''}(\omega)}{\hbar \omega},
\end{eqnarray}
with
\begin{eqnarray}
    P^{''}(\omega) 
    & \equiv &
    \frac{\pi^{2} \hbar^{2}}{4}
    \frac{1}{Z}
    \sum_{n,m}
    \left(
    e^{-\beta E_{m}} - e^{-\beta E_{n}}
    \right)  \nonumber \\
    &&
    \times
    \left| \langle n | \dot{N_{{\rm A}}} - \dot{N_{{\rm B}}} | m \rangle \right|^{2} \nonumber \\
    && \times
    \delta
    \left(
    \hbar \omega - ( E_{n}-E_{m} )
    \right).
\end{eqnarray}

We note that the operator $\dot{N_{\mu}} (\mu={\rm A,B})$ can be expressed by the localized operators
near the dots.
For example,
$\dot{N_{{\rm A}}} - \dot{N_{{\rm B}}}$ is given as follows;
\begin{eqnarray}
    \dot{N_{{\rm A}}}-\dot{N_{{\rm B}}}
    & = &
    \frac{2 {\rm i}}{\hbar} \sum_{k \sigma}
    V \left( \cos \frac{\varphi}{4} d_{{\rm e},\sigma}^{\dagger} \alpha_{{\rm a},k\sigma} \right. \nonumber \\
    && \left. +\sin \frac{\varphi}{4} d_{{\rm o},\sigma}^{\dagger} \alpha_{{\rm s},k\sigma} - h.c. \right).
\end{eqnarray}
In this place the operator
$\sum_{k\sigma} \alpha_{l,k\sigma} (l={\rm s,a})$,
which is given by linear combination of all $k$-state,
is proportional to the localized orbit near the dots.
Therefore the expression of conductance by using the quantity
$P^{''}(\omega)$ is suitable for the calculation based on the NRG method,
in which the approximate eigenstates 
are obtained successively starting from the localized orbits.

%

\clearpage

%

%


\begin{thebibliography}{99}



\bibitem{rf:Kondo_effect} See for example, A. C. Hewson:
{\it The Kondo problem to Heavy Fermions} 
(Cambridge University Press, Cambridge,1993) and references cited therein.



\bibitem{rf:Kondo_dot_Ng} T. K. Ng and P. A. Lee:
Phys. Rev. Lett. {\bf 61} (1988) 1768.
\bibitem{rf:Kondo_dot_JETP} L. I. Glazman and M. {\'E}. Ra{\u\i}kh:
JETP Lett. {\bf 47} (1988) 452.
\bibitem{rf:Kondo_dot_Kawa} A. Kawabata:
J. Phys. Soc. Jpn. {\bf 60} (1991) 3222.
\bibitem{rf:Oguri1} A. Oguri, H. Ishii and T. Saso:
Phys. Rev. B {\bf 51} (1995) 4715.



\bibitem{rf:Kondo_dot_Neq1} S. Hershfield, J. H. Davies and J. W. Wilkins:
Phys. Rev. Lett. {\bf 67} (1991) 3720.
\bibitem{rf:Kondo_dot_Neq2} Y. Meir, N. S. Wingreen and P. A. Lee:
Phys. Rev. Lett. {\bf 70} (1993) 2601.
\bibitem{rf:Kondo_dot_Neq3} A. L. Yeyati, A. Wart{\'\i}n-Rodero and F. Flores:
Phys. Rev. Lett. {\bf 71} (1993) 2991.
\bibitem{rf:Kondo_dot_Neq4} T. Inoshita, A. Shimizu, Y. Kuramoto and H. Sasaki:
Phys. Rev. B {\bf 48} (1993) 14725.



\bibitem{rf:And_dot_Eq0} L. I. Glazman and K. A. Matveev:
JETP Lett. {\bf 51} (1990) 485.
\bibitem{rf:And_dot_Eq1} Y. Meir, N. S. Wingreen and P. A. Lee:
Phys. Rev. Lett. {\bf 66} (1991) 3048.



\bibitem{rf:And_dot_Neq2} A. Groshev, T. Ivanov and V. Valtchinov:
Phys. Rev. Lett. {\bf 66} (1991) 1082.



\bibitem{rf:AB_dot_Akera} H. Akera:
Phys. Rev. B {\bf 47} (1993) 6835.



\bibitem{rf:AB_dot_Yacoby} A. Yacoby, M. Heiblum, D. Mahalu, and H. Shtrikman:
Phys. Rev. Lett. {\bf 74} (1995) 4047.
\bibitem{rf:AB_dot_Yacoby2} A. Yacoby, R. Schuster, and M. Heiblum:
Phys. Rev. B {\bf 53} (1996) 9583.



\bibitem{rf:AB_dot_Yeyati} A. L. Yeyati and M. B{\"u}ttiker:
Phys. Rev. B {\bf 52} (1995) 14360.
\bibitem{rf:AB_dot_Hachen} G. Hackenbroich and H. A. Weidenm{\"u}ller:
Phys. Rev. Lett. {\bf 76} (1996) 110.
\bibitem{rf:AB_dot_Bruder} C. Bruder, R. Fazio, and H. Shoeller:
Phys. Rev. Lett. {\bf 76} (1996) 114.



\bibitem{rf:NRG_Kri} H. R. Krishnamurthy, J. W. Wilkins ans K. G. Wilson:
Phys. Rev. B {\bf 21} (1980) 1003; {\bf 21} (1980) 1044.
\bibitem{rf:NRG_sakai} O. Sakai, S. Suzuki and Y. Shimizu: 
Physica B {\bf 206 \& 207} (1995) 141.
\bibitem{rf:NRG_shimizu} Y. Shimizu and O. Sakai:
{\it Computational Physics as a New Frontier in Condensed Matter Reseach} 
ed. H. Takayama, M. Tsukada, H. Shiba, F. Yonezawa, M. Imada and Y. Okabe
(The Physical Society of Japan, 1995) p.42.



\bibitem{rf:linear} R. Kubo:
J. Phys. Soc. Jpn. {\bf 12} (1957) 570.



\bibitem{rf:Friedel_L} D. C. Langreth:
Phys. Rev. {\bf 150} (1966) 516.
%
\bibitem{rf:Friedel} H. Shiba:
Prog. Theor. Phys. {\bf 54} (1975) 967.



%
\bibitem{rf:n_e=n_o} 
In general,
there is no reason to be 
$\langle n_{0,{\rm e}} \rangle = \langle n_{0,{\rm o}} \rangle =1$.
%
This result is critically dependent on the assumption that 
the hybridization of each channel has electron-hole symmetry.
%
In realistic cases,
AB oscillations with small amplitude are expected.


%
\bibitem{rf:-cos}
For the temperatures higher than $T_{\rm K}$,
and in the parameters region 
$\varepsilon_{\rm d} < \varepsilon_{\rm F} < \varepsilon_{\rm d} + U$,
one can find that the conductance is proportional to
$(K^{2} + 4 J^{2} \langle \vec{S_{1}} \cdot \vec{S_{2}} \rangle) \cos\varphi + C$,
where $K=(1/\varepsilon_{\rm d} + 1/(\varepsilon_{\rm d}+U) )$,
$J=(1/\varepsilon_{\rm d} - 1/(\varepsilon_{\rm d}+U) )$
and $C$ is constant in $\varphi$.
%
(This expression is obtained
following Akera's consideration.~\cite{rf:AB_dot_Akera})
%
Usually we can expect $4 J^{2} \gg K^{2}$
and $\langle \vec{S_{1}} \cdot \vec{S_{2}} \rangle < 0$,
so the coefficient of $\cos\varphi$ could have the negative sign.
%
The numerical results in $T \ge T_{11}$ 
might be shown in such a case.
%



\end{thebibliography}
\end{document}